\documentclass[preprint,aps,floatfix,superscriptaddress,pre,showpacs]{revtex4}
\usepackage{amsmath, amsthm, amssymb, graphicx}
\input epsf.sty
\begin{document}

\def\g{\gamma}
\def\r{\rho}
\def\w{\omega}
\def\wo{\w_0}
\def\wp{\w_+}
\def\wm{\w_-}
\def\t{\tau}
\def\av#1{\langle#1\rangle}
\def\pf{P_{\rm F}}
\def\pr{P_{\rm R}}
\def\F#1{{\cal F}\left[#1\right]}

\title{Interplay between partnership formation and competition in generalized May-Leonard games}

\author{Ahmed Roman}
\affiliation{Department of Physics, Virginia Polytechnic Institute and State University, Blacksburg, Virginia 24061-0435, USA}

\author{Debanjan Dasgupta}
\affiliation{Department of Physics, University of Virginia, Charlottesville, Virginia 22904-4133, USA }

\author{Michel Pleimling}
\affiliation{Department of Physics, Virginia Polytechnic Institute and State University, Blacksburg, Virginia 24061-0435, USA}

\date{\today}

\begin{abstract}
In order to better understand the interplay of partnership and competition 
in population dynamics we study a family of generalized
May-Leonard models with $N$ species. These models have a very rich
structure, characterized by different types of space-time patterns. Interesting partnership
formations emerge following the maxim that "the enemy of my enemy is my friend." In specific cases
cyclic dominance within coarsening clusters yields a peculiar coarsening behavior with intriguing pattern
formation. We classify the different types of dynamics through the analysis of the square of the
adjacency matrix. The dependence of the population densities on 
emerging pattern and propagating wave fronts
is elucidated through a Fourier analysis. Finally, after having identified collaborating teams,
we study interface fluctuations where we initially populate different parts of the system with different teams.
\end{abstract}

\maketitle

\section{Introduction}
In many ecological as well as biological systems biodiversity is realized despite the competition between
different species for limited resources. It is therefore a fundamental problem in ecology 
to understand the mechanisms underlying biodiversity and its stability over time \cite{May74,Smith74,Sole06}.
Models of predator-prey interactions, in which the survival of a species heavily depends on other 
species' behavior, provide reliable means for studying the emergence of biodiversity. Interestingly,
these models have a direct relationship with evolutionary game theory \cite{Smith82,Hof98,Now06,Sza07,Fre09}. 
In the last few years systems with simple cyclic dominance of competing species have attracted wide interest as they 
provide a platform for investigating species diversity. Most of these studies focused on cases
involving three \cite{Fra96a,Fra96b,Pro99,Tse01,
Ker02,Kir04,Rei06,Rei07,Rei08,Cla08,Pel08,Rei08a,Ber09,Ven10,Shi10,
And10,Wan10,Mob10,He10,Win10,He11,Rul11,Wan11,Nah11,Jia11,Pla11,Dem11,He12,Van12,Don12,Dob12,
Juu12,Lam12,Jia12,Ada12,Juu12a} 
or four species \cite{Fra96a,Fra96b,Kob97,Sat02,Sza04,He05,Sza07b,Sza08,Cas10,
Dur11,Dur12,Rom12,Dob12} in cyclic competition. 
Thereby the emergence of spatio-temporal patterns in lattice systems,
where fluctuations and correlations induce species co-existence, received special attention.
More complex situations, involving either more species and/or more complicated interactions than simple
cyclic competition, have been the subject of rather few recent studies \cite{Fra96a,Fra96b,Kob97,
Fra98,Sza01,Sza01b,Sat02,He05,Sza05,Sza07b,Sza07c,Per07,Sza08,Sza08a,Nob11,Zia11,Lut12,Van12,Ave12a,Ave12b}.
However, these more complex cases are obviously closer to realistic situations where in a given
ecological environment multiple species are involved in a complicated food and domination web.

The May-Leonard model \cite{May75} and its spatial stochastic version \cite{Rei07} have yielded
many important insights into pattern formation in the presence of cyclic competition. However, for more
involved cases emerging spatio-temporal patterns as well as complicated coarsening processes
have yet to be studied systematically. In two recent papers \cite{Ave12a,Ave12b} Avelino {\it et al}
introduced a set of models with $N$ species that possess a $Z_N$ symmetry and discussed the
appearance of intriguing pattern in two-dimensional lattice systems. In the present paper we discuss a related 
family of models composed of $N$ species and elucidate some of their characteristic properties. Our aim
is thereby to gain a better understanding of the complicated interplay between partnership and competition
in a spatial environment occupied by more than three species.

Our paper is organized in the following way. In Section II we introduce our generalization of the
stochastic May-Leonard model, whereas in Section III we discuss a few selected cases that illustrate
the very rich coarsening behaviors and spiral formations
that can show up in this type of system. Indeed, depending on the
number of species and the chosen interaction schemes, very different spatio-temporal behaviors emerge,
ranging from simple coarsening of compact domains to propagating wave fronts.
More complex situations involve global competition between different alliances where
members of each alliance prey on each other, yielding an intriguing mixture of coarsening process 
and wave front propagation. Based on an analysis of the
square of the adjacency matrix, we are able to predict the dynamics that is
realized in a given system. In Sections IV and V we present a more quantitative study of
some of the aspects related to pattern formation. Whereas in Section IV we quantify 
through a temporal Fourier analysis the dependence
of particle densities on system parameters,
in Section V we analyze the time dependence of growth lengths and interface fluctuations for cases where
coarsening processes take place. Finally, in Section VI we discuss our results and conclude.

\section{A family of generalized May-Leonard models}
We consider in the following individuals of $N$ different species that live on a two-dimensional lattice. 
Each individual is mobile, can prey on individuals from some of the other species, and can give birth 
to off-springs. The events that are possible at each instant are determined by the local environment of the
individual under consideration. Denoting by $X_i$ a member of species $i$ sitting at a selected site, 
we can for the most general
case capture the different events in the form of chemical reactions:
\begin{eqnarray}
X_{i} + {\emptyset}& \xrightarrow{\beta_{i}} & {\emptyset}+X_{i} \label{model1} \\
X_{i} + {\emptyset}& \xrightarrow{\gamma_{i}} & X_{i}+X_{i} \label{model2} \\
X_{i} + X_{j}& \xrightarrow{\alpha_{ij}} & X_{j}+X_{i} \label{model3} \\
X_{i} + X_{j}& \xrightarrow{\delta_{ij}} & {\emptyset}+X_{i}\label{model4}  
\end{eqnarray}
The first two equations summarize the events when the selected neighboring site is unoccupied, as
indicated by the symbol $\emptyset$. In that case our individual can either jump to the empty site with
rate $\beta_{i}$ or an off-spring is deposited on the empty site with rate $\gamma_{i}$.
The last two equations, on the other hand, indicate the possible interactions if a member of
species $j$ sits on the neighboring site: the two individuals then either can swap places with
rate $\alpha_{ij}$ or, if species $j$ is a prey of species $i$, the individual $X_{j}$ is removed 
from the neighboring site with rate $\delta_{ij}$ and at the same time $X_i$ jumps to the 
newly unoccupied site.

A modified version of the model can be obtained by imposing that at every moment every site is occupied
by exactly one individual. Due to the absence of empty sites, the reactions then simplify to
\begin{eqnarray}
X_{i} + X_{j}& \xrightarrow{\alpha_{ij}} & X_{j}+X_{i} \label{model_singlesite1} \\
X_{i} + X_{j}& \xrightarrow{\delta_{ij}} & X_{i}+X_{i} \label{model_singlesite2}
\end{eqnarray}
so that predation and birth of an off-spring take place at the same time, with the mobility being reduced to
the swapping of individuals on neighboring sites. 

\begin{figure} [h]
\includegraphics[width=0.28\columnwidth]{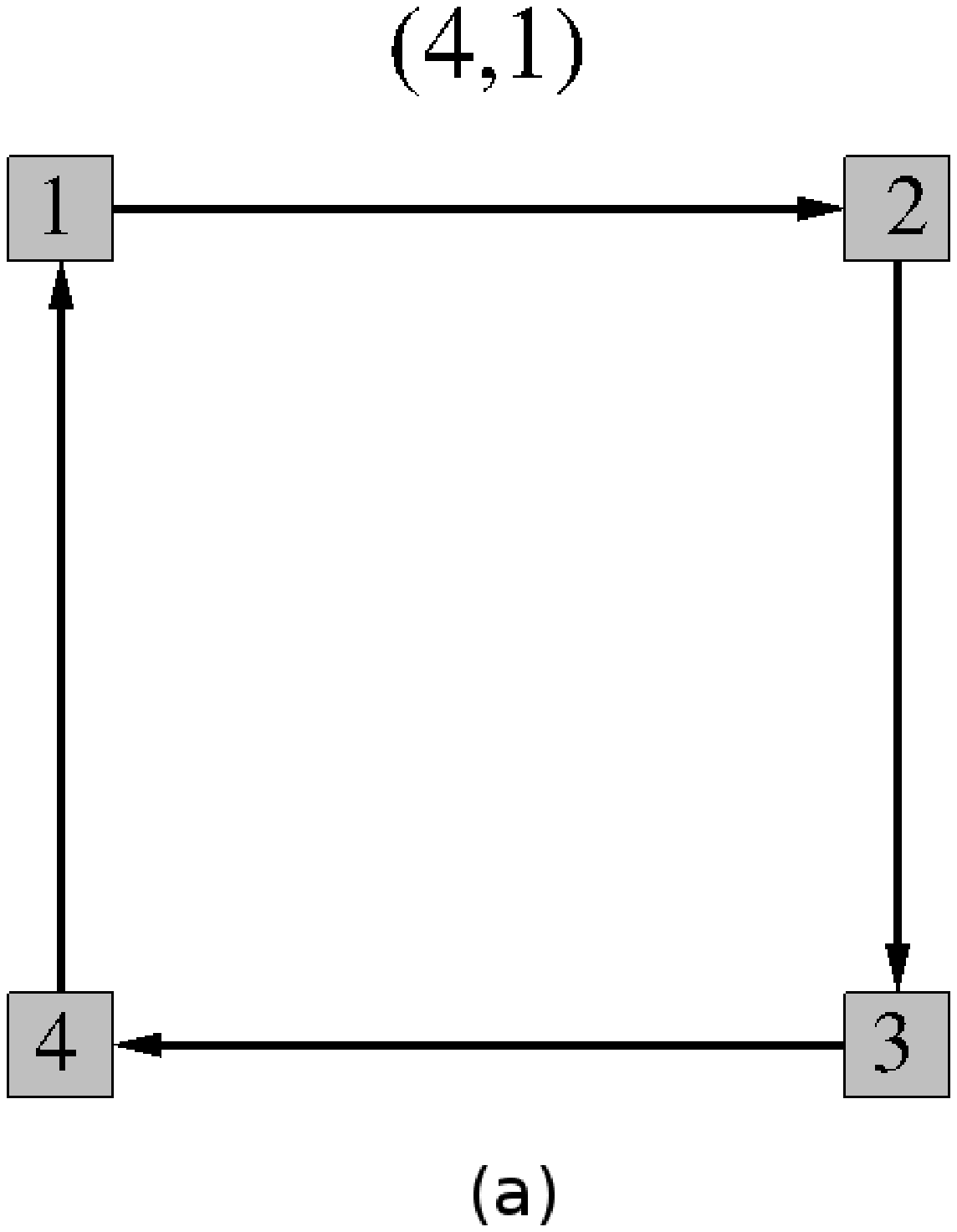}~~
\includegraphics[width=0.28\columnwidth]{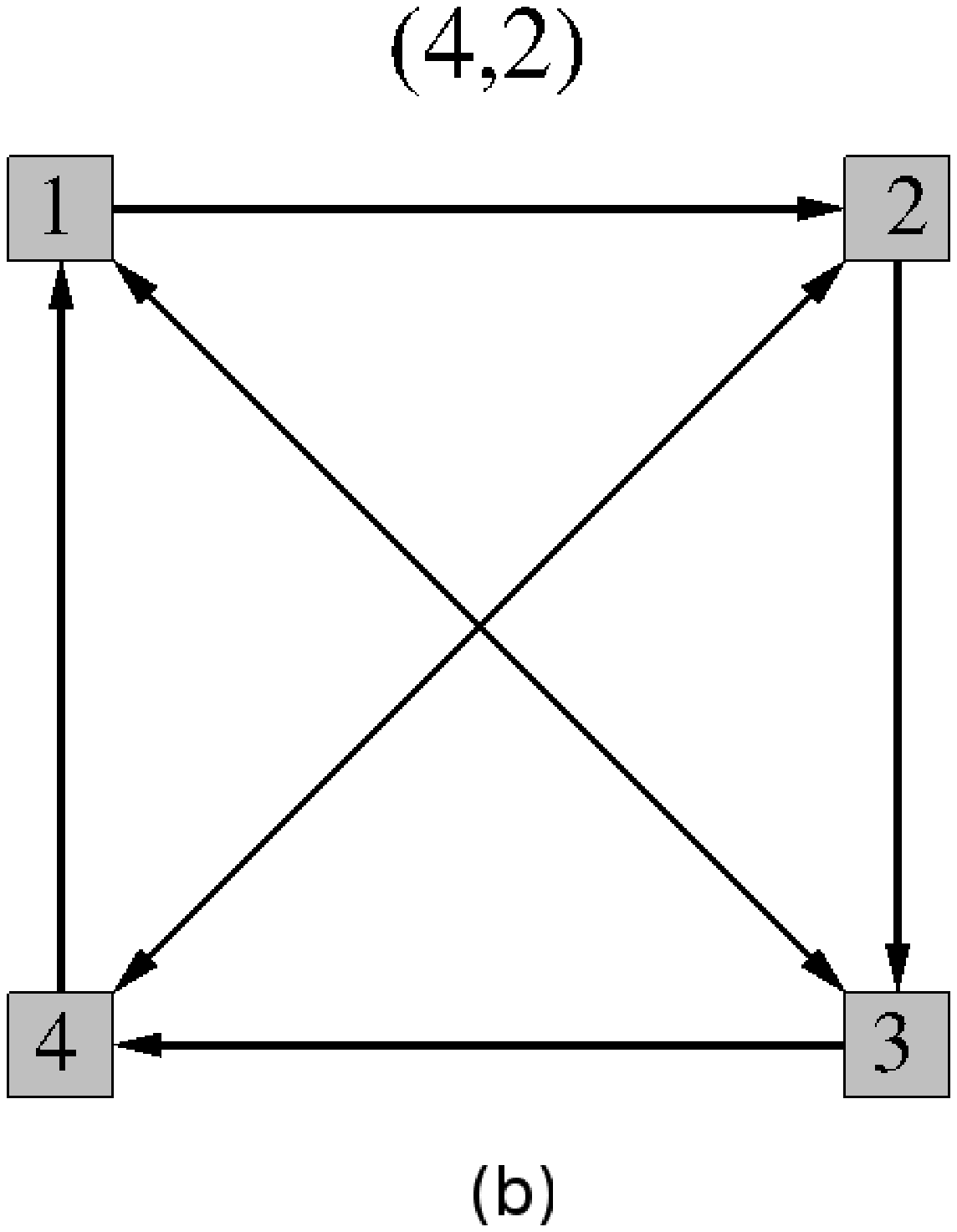}~~
\includegraphics[width=0.28\columnwidth]{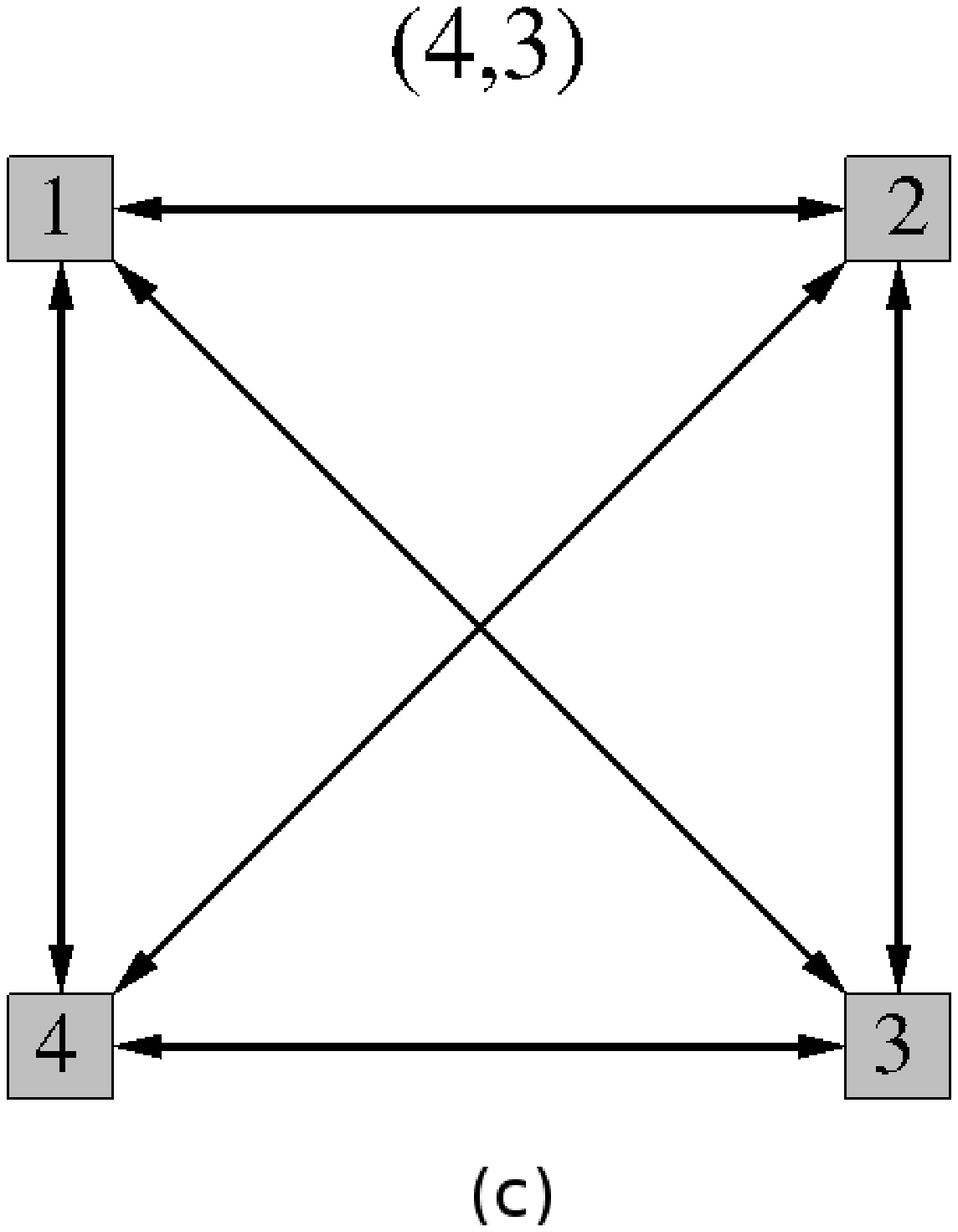}
\caption{The possible reaction schemes for a system with four species. The label $(N,r)$
indicates the number of species $N$ and the number of preys $r$ for every species.\label{fig1} 
}
\end{figure}

In the following we study a specific subclass of the very general model described by equations (\ref{model1})-(\ref{model4})
(we will also briefly look at the corresponding simplified version given by equations (\ref{model_singlesite1})-(\ref{model_singlesite2})
at the very end when discussing interface fluctuations).
We call model $(N,r)$, $r < N$, the generalized May-Leonard model with $N$ species where each species preys on
$r$ other species.
This is done in a cyclic way, i.e. species $i$ preys on species $i+1$, $i+2$, $\cdots$,
$i+r$ (this has to be understood modulo $N$). $(N,1)$ is therefore identical to the usual
$N$-species rock-paper-scissors game where every species preys on a single species and at the
same time is the prey of a different unique predator. Fig. \ref{fig1} illustrates the different predation
schemes that are possible for the simple case of four species.

As we will see in the next Section, in many cases the characteristic behavior of the systems, ranging from
simple coarsening to wave propagation, can be read off directly from the square of the adjacency matrix $\underline{A}$.
For every model the graph of the predation interaction, see the examples in Fig. \ref{fig1}, can be represented
by the $N \times N$ adjacency matrix, where the element $A_{ij} = 1$ if there is a directed edge from
$i$ to $j$ and zero otherwise. As we do not allow that members of a given species prey on themselves, the
adjacency matrix always contains zeros on the diagonal. As a simple example consider the case $(4,2)$,
see the middle panel in Fig. \ref{fig1}. From the directed edges we immediately obtain that
\begin{equation}
\renewcommand{\arraystretch}{.45}
\underline{A} = \left( \begin{array}{cccc}
0 & 1 & 1 & 0 \\
0 & 0 & 1 & 1 \\
1 & 0 & 0 & 1 \\
1 & 1 & 0 & 0
\end{array} \right)~.
\end{equation}
Most importantly, the square of the adjacency matrix $\underline{B} = \underline{A}^2$ contains information about preferred
partnership formations. The element $b_{ij}$ counts the number of directed paths of length 2 from vertex $i$
to vertex $j$, i.e. the number of paths of the form $i \longrightarrow k \longrightarrow j$ where $k$ is a vertex
different from $i$ and $j$. Species $j$ then has a preference to ally with the species that preys on most of 
its own predators, following the maxim that "the enemy of my enemy is my friend." This preferred ally of species
$j$ is directly obtained from the condition $\max\limits_i b_{ij}$. In the next Section we will discuss a range of
non-trivial examples that illustrate the importance of the matrix $\underline{B}$.

\section{Partnership and competition}
We first focus in the following on some cases with six and nine species. This allows us to discuss the typical 
scenarios and to illustrate how the matrix $\underline{B}$ can be used to predict the dynamics of the system.
We end this Section with a discussion of expected scenarios for any number of species.

In our simulations we first select a site and then choose one of the neighboring sites of this 
first site. Depending on what species occupy the different sites, different
interactions are possible, as described in Section II. In general, the individual $X_i$ sitting on the first selected
site has precedence. If the individual $X_j$ sitting on the second site is both prey and predator of $X_i$, it is $X_i$ that is allowed
to attack $X_j$ or, if that fails, to swap places with $X_j$. If $X_i$ is not a predator
of $X_j$, then the latter is allowed to attack the former.

For all the simulations reported in this Section we consider systems composed of $400 \times 400$ lattice sites. In the
initial state every lattice site is either empty or populated with a member of any of the $N$ species with the 
same probability $1/(N+1)$. 
The number of empty sites rapidly decreases in the early stages of the simulation as every
species {\it feeds} on the empty sites, thereby approaching rapidly
a steady state value that depends on the chosen reaction scheme as
well as on the values of the reaction constants. Empty sites can not survive 
inside ordered domains because they are used by all species in the birth process.
The only way to generate new empty sites is through predation between different species.
This process only occurs near the boundaries separating different domains or spiral arms.
All snapshots have been taken after 500 sweeps,
where one sweep corresponds to selecting on average every lattice site once. 
For the predation and birth events the rates have been set equal to 0.2, whereas for the exchanges between two individuals
and the jumps of an individual to an empty site the rates have been fixed at 0.16 for interacting species and at 0.2 for
non-interacting species. The emerging spatio-temporal structures are independent of the specific values of these rates,
as long as these rates are the same for all species and are not zero.

\subsection{Six species}

\begin{figure} [h]
\includegraphics[width=0.48\columnwidth]{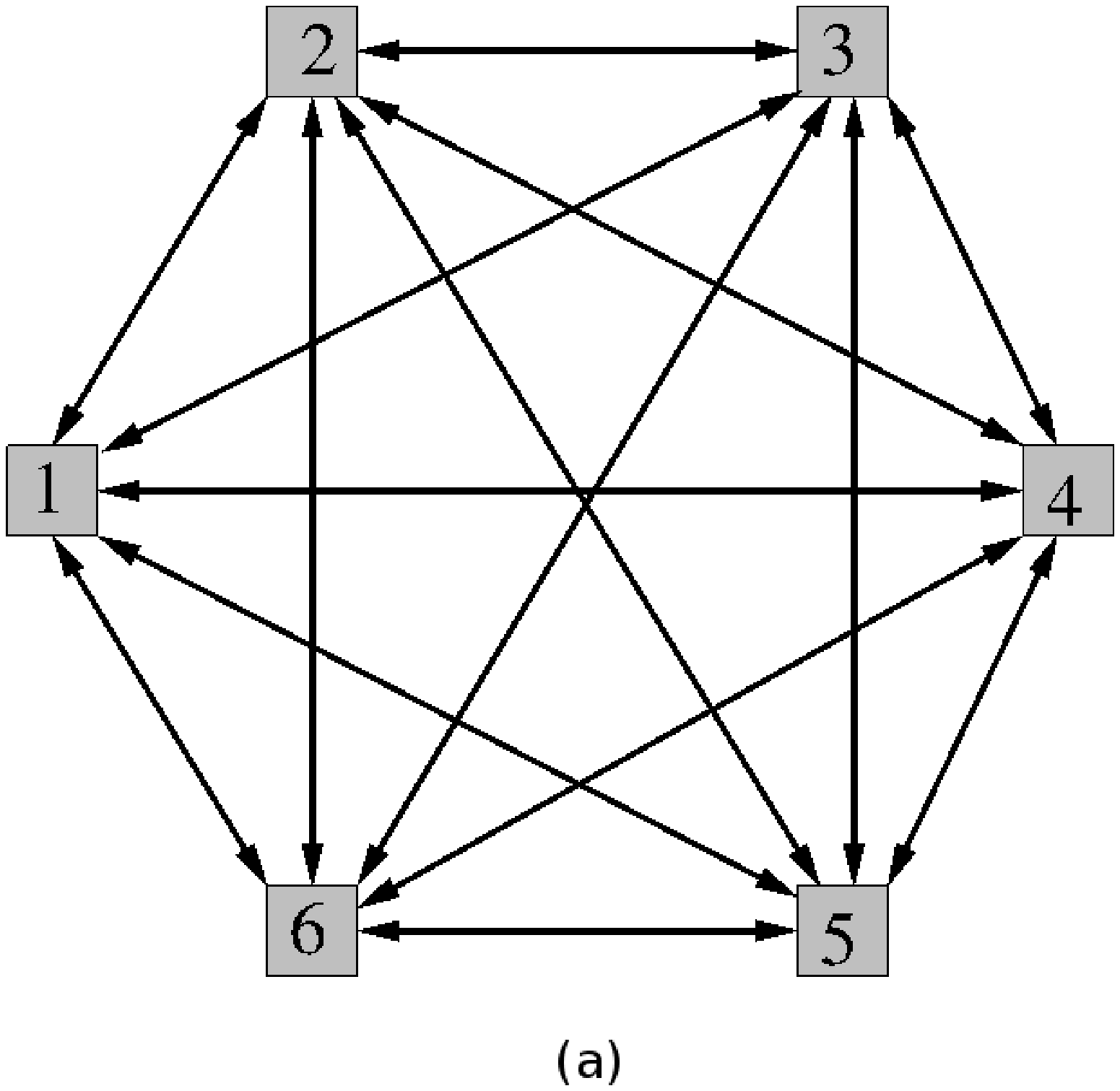}~~
\includegraphics[width=0.48\columnwidth]{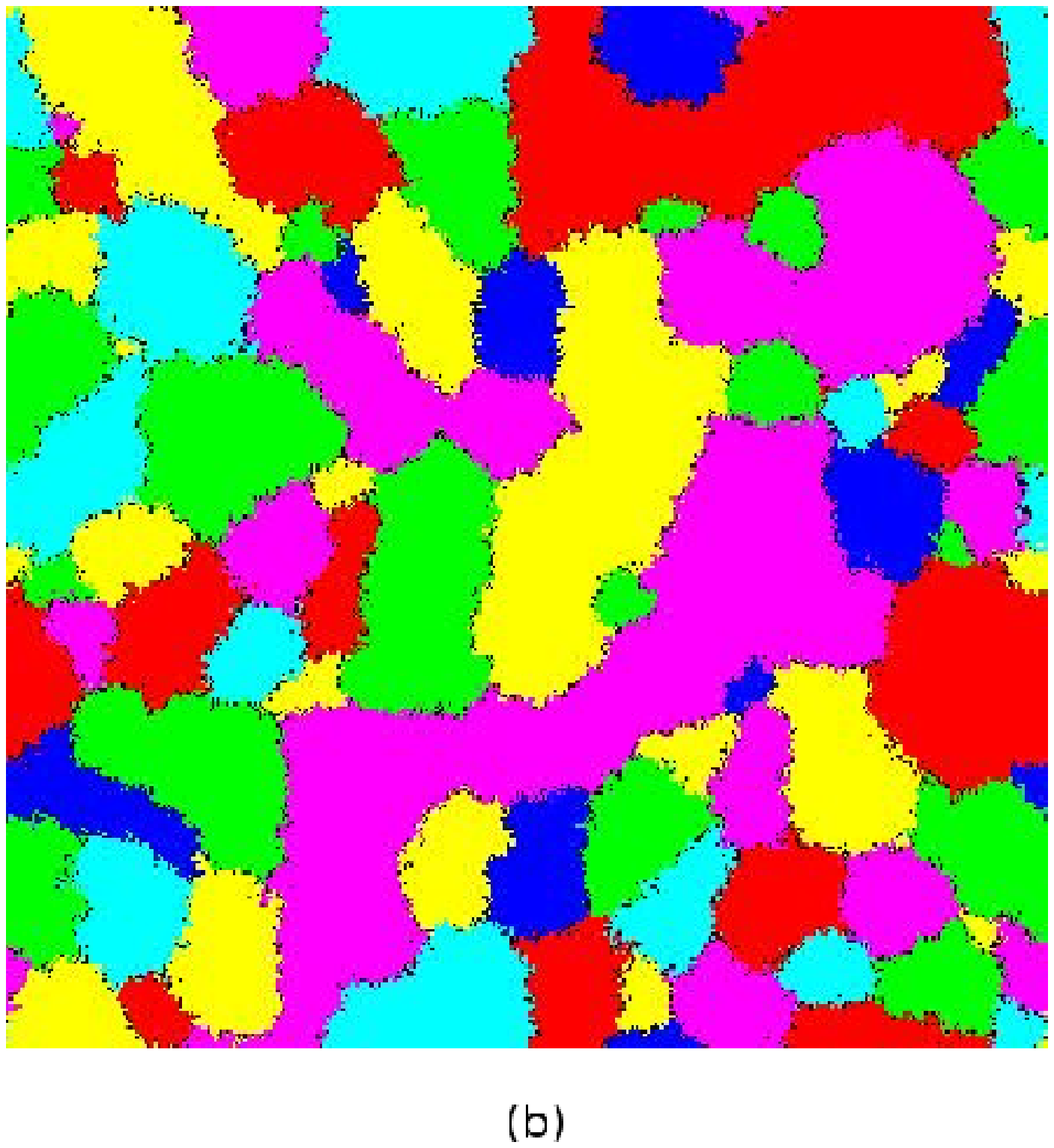}~~
\caption{(Color online) In the model $(6,5)$ every species preys on every other species. As a result the 
members of each species form compact clusters, as this increases their chances of survival.
\label{fig2}
}
\end{figure}

We start with the extreme case $(6,5)$ where every species preys on every other, see the directed graph in Fig.\ \ref{fig2}.
The corresponding adjacency matrix has entries 1 everywhere except for the zero diagonal elements.
The square of that matrix is then given by
\begin{equation}
\renewcommand{\arraystretch}{.45}
\underline{B} = \left(
\begin{array}{cccccc}
 5 & 4 & 4 & 4 & 4 & 4 \\
 4 & 5 & 4 & 4 & 4 & 4 \\
 4 & 4 & 5 & 4 & 4 & 4 \\
 4 & 4 & 4 & 5 & 4 & 4 \\
 4 & 4 & 4 & 4 & 5 & 4 \\
 4 & 4 & 4 & 4 & 4 & 5
\end{array}
\right)~.
\end{equation}
For every column $j$ the maximum value is found at $i = j$: the species $j$ does not want to ally itself with any of the other species, as all
other species prey on $j$. Consequently, the members of species $j$ cluster around each other, see the snapshot
in Fig.\ \ref{fig2}. We then have the typical situation of a coarsening process with six competing states, where smaller domains
vanish at the expense of the larger domains.
The domains are thereby very compact. 
This is very reminiscent of the coarsening process in a six-state Potts model \cite{Lou10,Lou12},
especially at low temperatures where thermal fluctuations are small. In fact, we checked that the typical domain size
increases as a square-root of time, as it is expected for a curvature-driven coarsening process.

\begin{figure} [h]
\includegraphics[width=0.48\columnwidth]{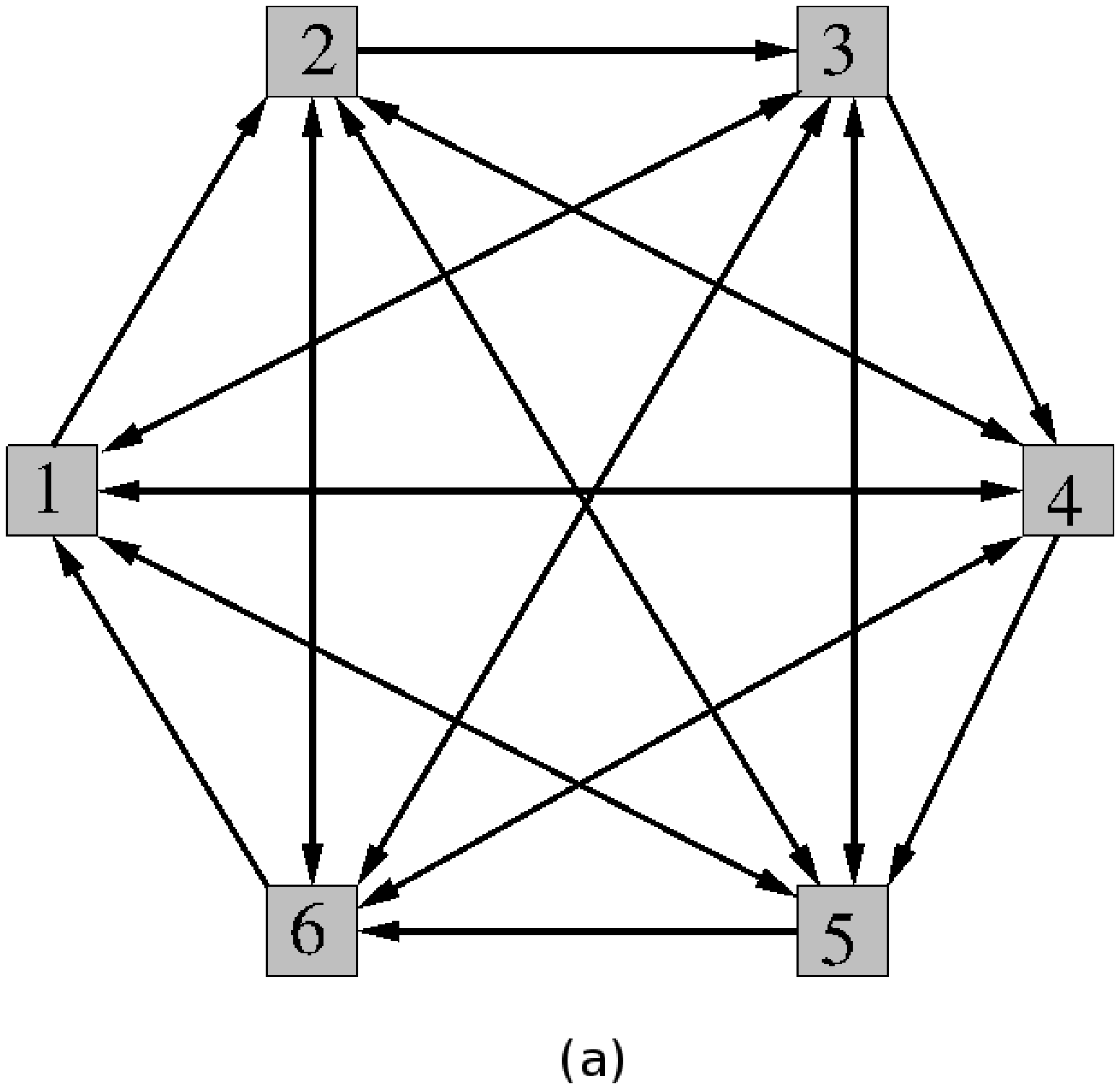}~~
\includegraphics[width=0.48\columnwidth]{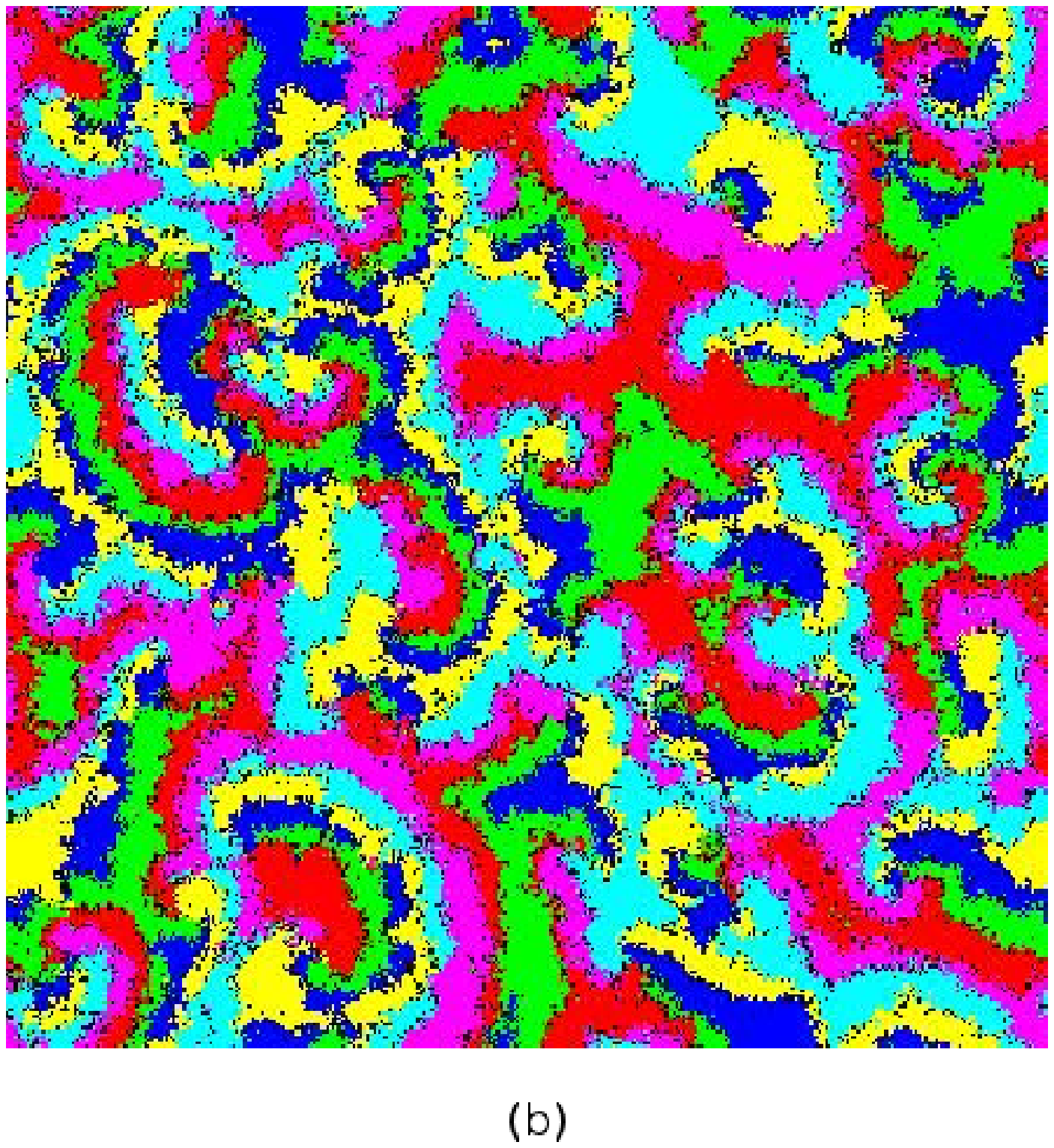}~~
\caption{(Color online) The model $(6,4)$ is characterized by propagating wave fronts, where the order of the wave fronts
can be read off from the square of the adjacency matrix.
\label{fig3}
}
\end{figure}

Model $(6,4)$ is characterized by the fact that for every species there is exactly one prey which is not
at the same time predator of that species. In addition this prey is feeding upon all four predators
of the species under consideration. As a direct illustration of the maxim that "the enemy of my enemy is my
friend," our species therefore prefers to closely follow this prey as this increases its chances of survival.
This behavior is nicely captured by the square of the adjacency matrix
\begin{equation}
\renewcommand{\arraystretch}{.45}
\underline{B} = \left(
\begin{array}{cccccc}
 3 & 2 & 2 & 2 & 3 & 4 \\
 4 & 3 & 2 & 2 & 2 & 3 \\
 3 & 4 & 3 & 2 & 2 & 2 \\
 2 & 3 & 4 & 3 & 2 & 2 \\
 2 & 2 & 3 & 4 & 3 & 2 \\
 2 & 2 & 2 & 3 & 4 & 3
\end{array}
\right)~.
\end{equation}
From the first column follows that species 1 wants to ally itself with species 2, the only one of its prey
which is not at the same time preying on 1, which itself would like to
stay close to species 3 etc. Hence the species try to make alliances in a cyclic way, which 
gives raise to the spatio-temporal pattern shown in Fig.\ \ref{fig3}. The order of the propagating wave fronts
is thereby fixed by the matrix $\underline{B}$, i.e. species $i$ follows species $i+1$, which follows species $i+2$, and so on.
These wave fronts are created in different parts of the systems, which leads to the shown complex pattern with
propagating and annihilating wave fronts.

\begin{figure} [h]
\includegraphics[width=0.48\columnwidth]{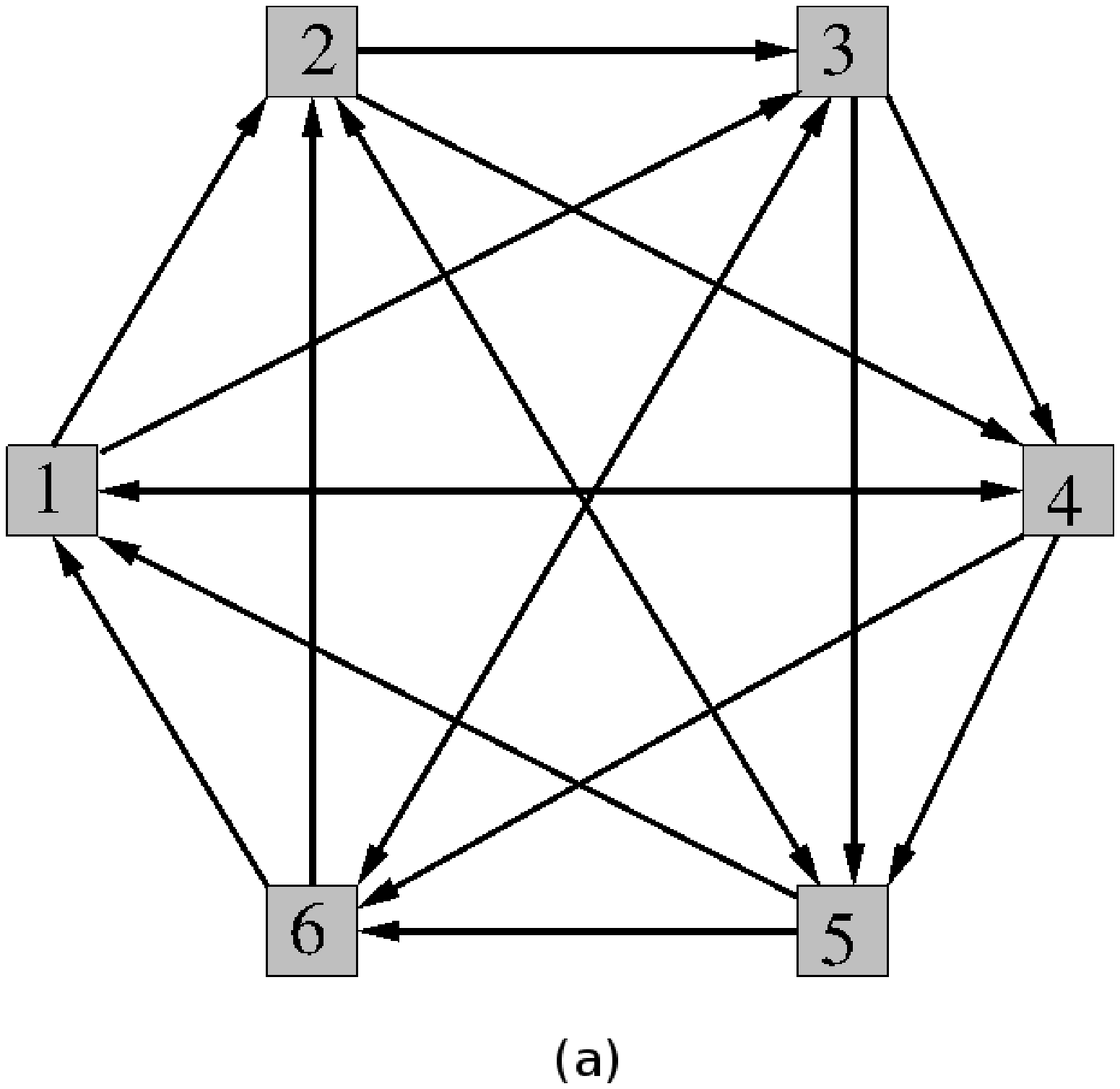}~~
\includegraphics[width=0.48\columnwidth]{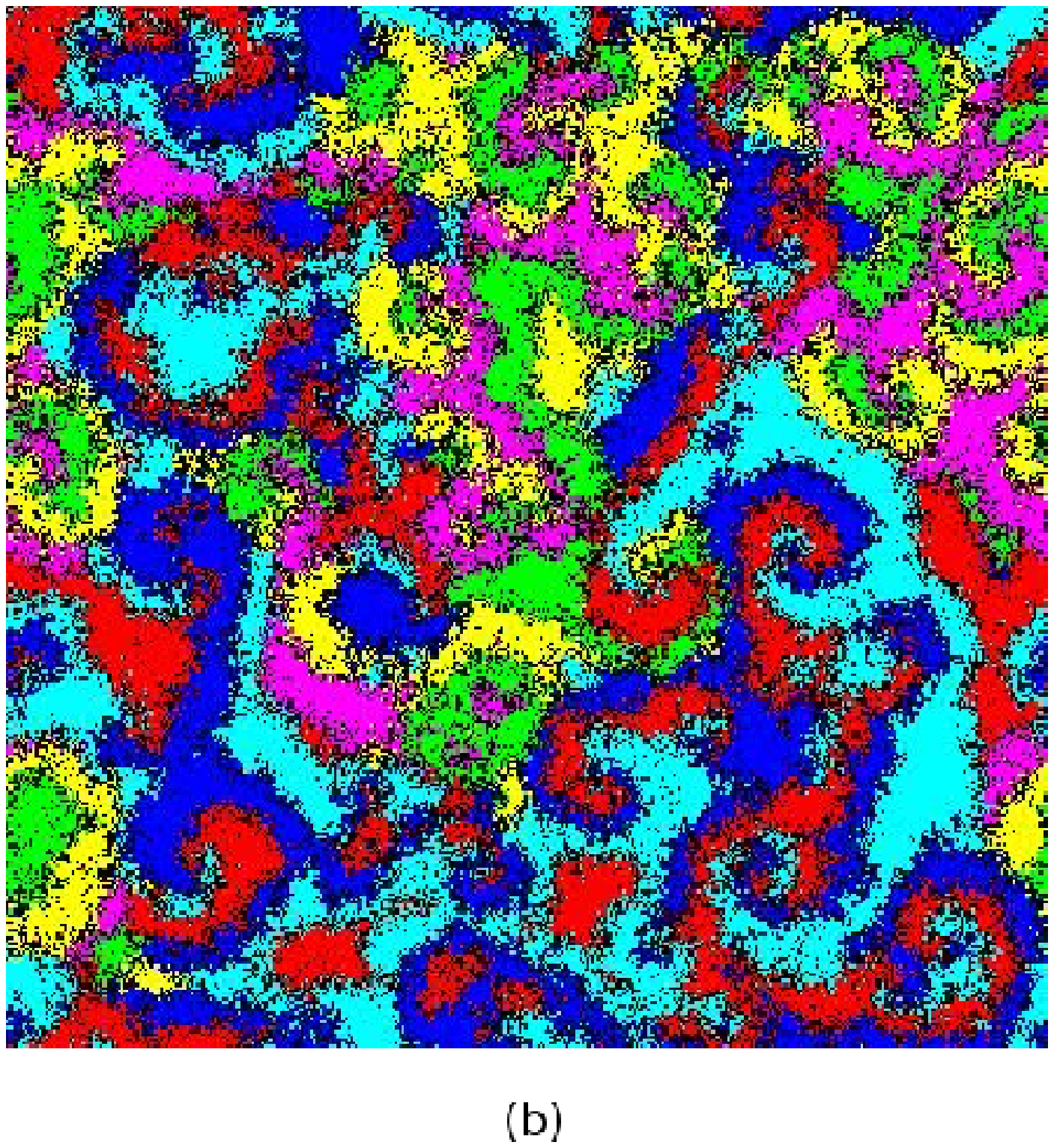}~~
\caption{(Color online) In the model $(6,3)$ two different types of domains coarsen. Within each domain
three species undergo a cyclic rock-paper-scissors game.
\label{fig4}
}
\end{figure}

The case $(6,3)$ is a very interesting one as it combines coarsening of domains with wave fronts {\it inside}
the domains, see Fig. \ref{fig4}. Inspection of the matrix
\begin{equation}
\underline{B} =
\renewcommand{\arraystretch}{.45}
\left(
\begin{array}{cccccc}
 1 & 0 & 1 & 2 & 3 & 2 \\
 2 & 1 & 0 & 1 & 2 & 3 \\
 3 & 2 & 1 & 0 & 1 & 2 \\
 2 & 3 & 2 & 1 & 0 & 1 \\
 1 & 2 & 3 & 2 & 1 & 0 \\
 0 & 1 & 2 & 3 & 2 & 1
\end{array}
\right)
\end{equation}
reveals the formation of two alliances, namely [1,3,5] and [2,4,6]. The species 1, 3, and 5 join forces in order
to fight off  the remaining
species which are perceived to be the greater danger. This then yields a competition between two different domain types.
However, within the domains, each species is paired up with both one of its prey and one of its predators. As a
result a cyclic rock-paper-scissors game sets in within each domain, yielding an intriguing combination of
domain coarsening and non-trivial internal dynamics. Whereas for the case (6,5) discussed above 
smaller domains vanish because individuals are eaten up by the surrounding predators, in the case (6,3)
the probable scenario is that inside a domain one of the predators is too successful and causes the extinction
of its prey. If that happens then the remaining two species are also vanishing rapidly as they
can not resist any more the pressure of the surrounding predators. It is therefore mandatory for the survival
of the species inside each domain that they arrange themselves in a predator-prey relationship that is moderate
enough so that no species is completely eaten up by its ally. This scenario can be captured by the maxim that
inside the domain "every species must do what is best for both itself and the alliance."

\begin{figure} [h]
\includegraphics[width=0.48\columnwidth]{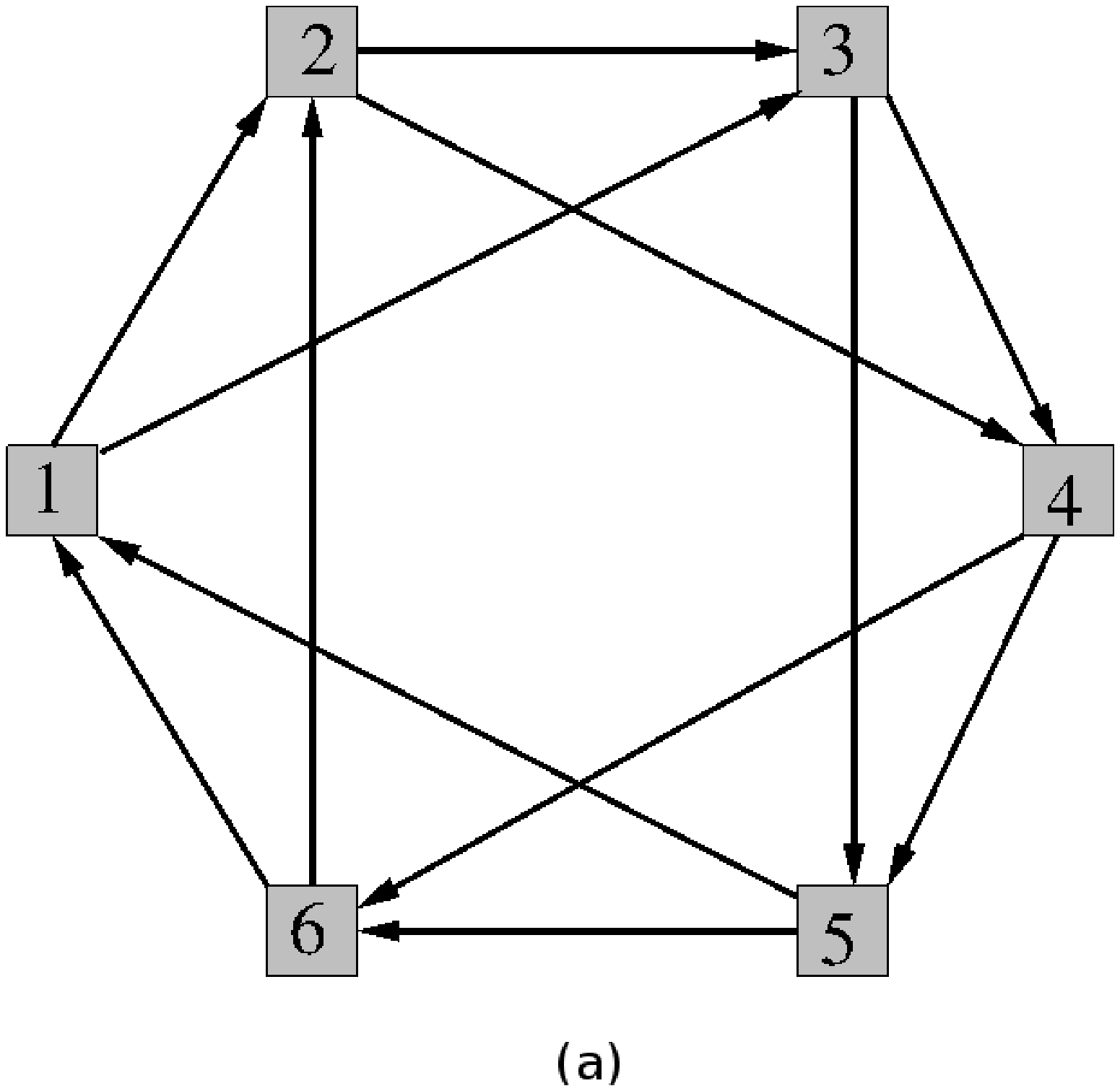}~~
\includegraphics[width=0.48\columnwidth]{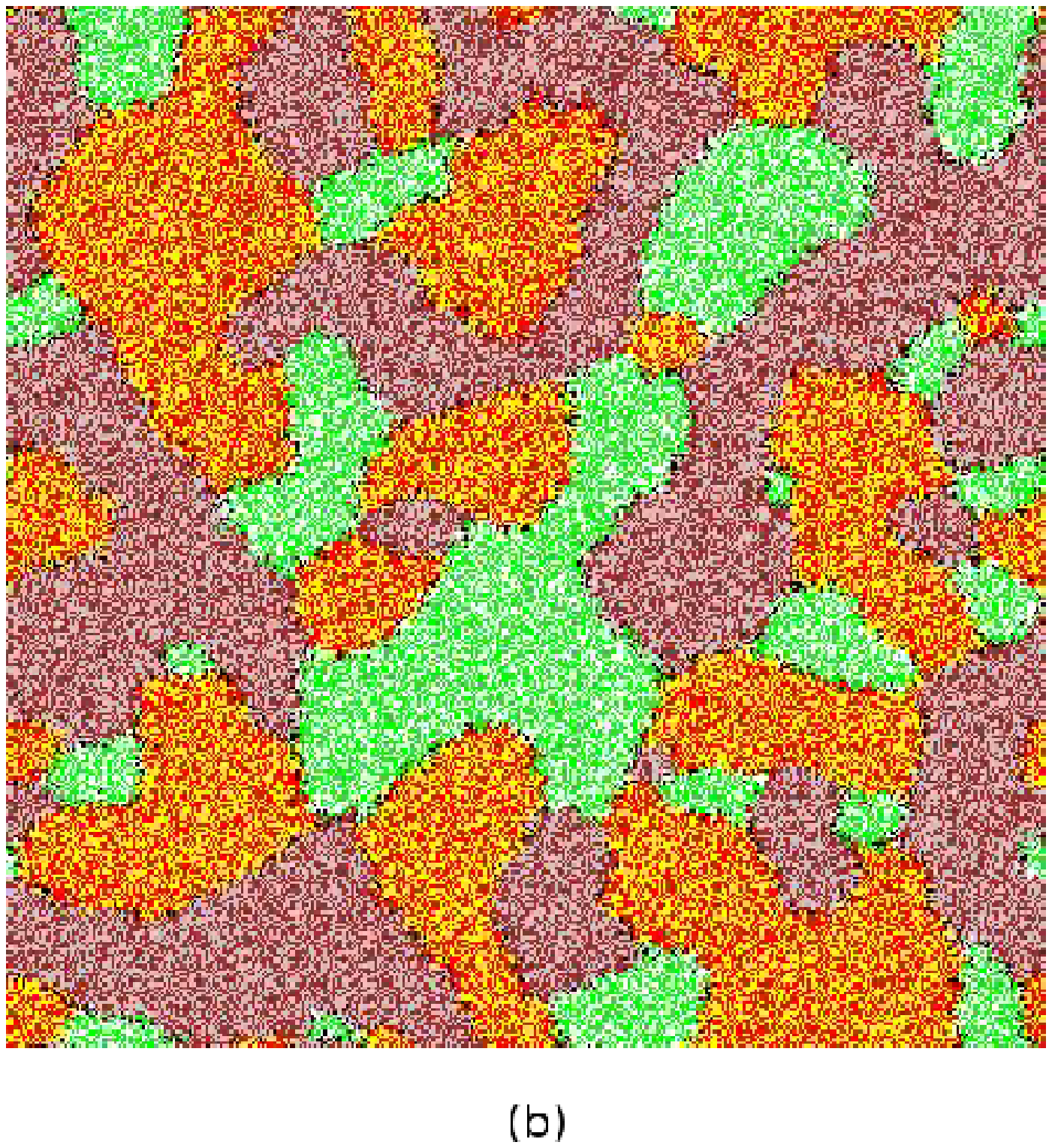}~~
\caption{(Color online) Model $(6,2)$ is characterized by the emergence of three competing teams where each
team is formed by two neutral, i.e. non-interacting, species. Note that for clarity we use here different colors
than for the other figures with six species.
\label{fig5}
}
\end{figure}

In the model (6,2) we have the situation that not all species are interacting. As a consequence, neutral, i.e. non-interacting,
species tend to come together, thereby exploiting the protection they receive from their partner. This
again immediately follows from the matrix
\begin{equation}
\underline{B} =
\renewcommand{\arraystretch}{.45}
\left(
\begin{array}{cccccc}
 0 & 0 & 1 & 2 & 1 & 0 \\
 0 & 0 & 0 & 1 & 2 & 1 \\
 1 & 0 & 0 & 0 & 1 & 2 \\
 2 & 1 & 0 & 0 & 0 & 1 \\
 1 & 2 & 1 & 0 & 0 & 0 \\
 0 & 1 & 2 & 1 & 0 & 0
\end{array}
\right)
\end{equation}
which implies the formation of three teams [1,4], [2,5], and [3,6]. Consequently, we observe the formation and coarsening
of three types of domains, each containing two mutually neutral species, see Fig. \ref{fig5}. The dynamics therefore is
expected to be similar to that of the three states Potts model. Alliance formation and domain growth have also been observed
previously in related six-species models where every species has two preys and at the same time is the prey of two other
species \cite{Sza01b,Sza05}.

\begin{figure} [h]
\includegraphics[width=0.48\columnwidth]{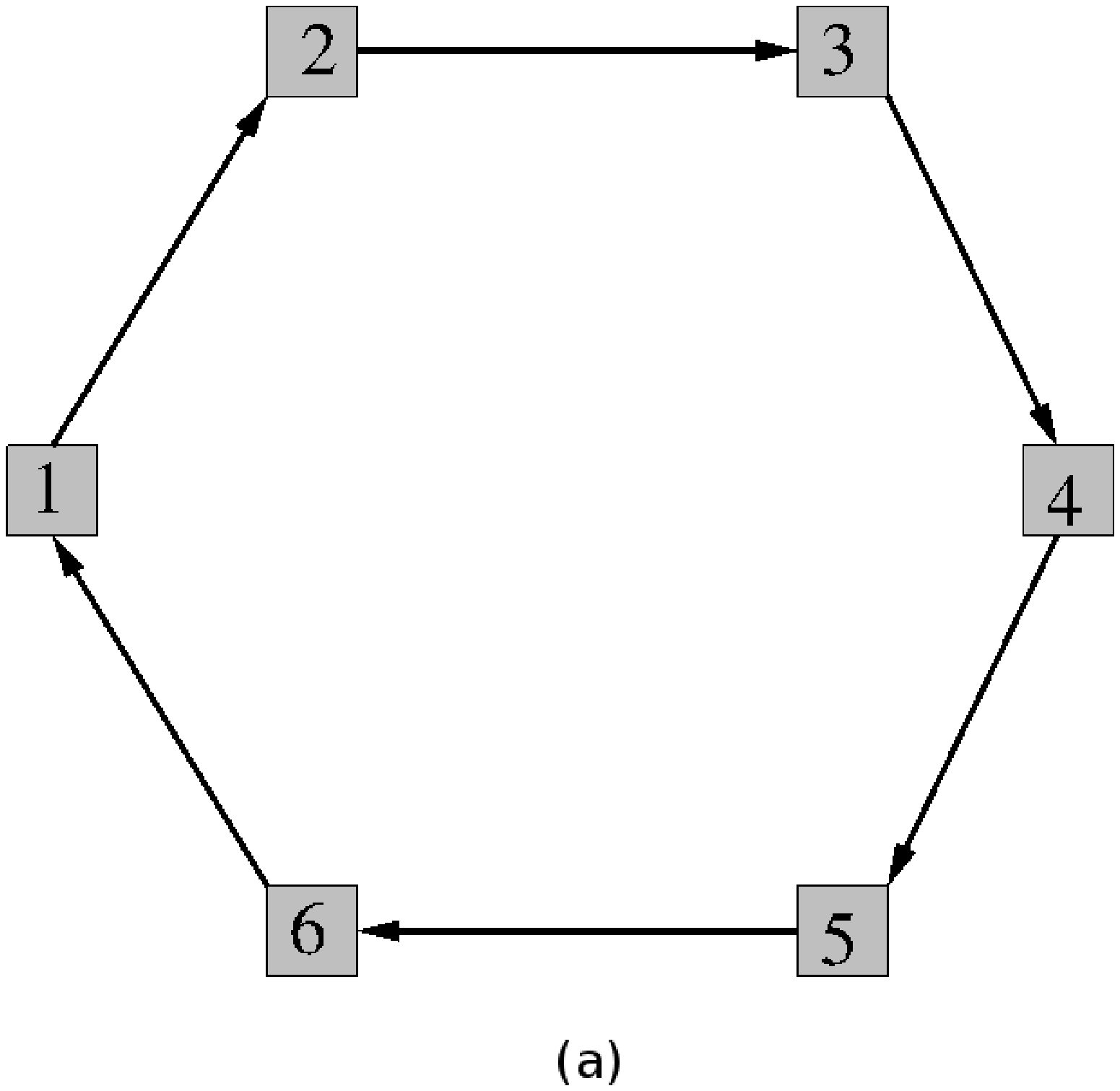}~~
\includegraphics[width=0.48\columnwidth]{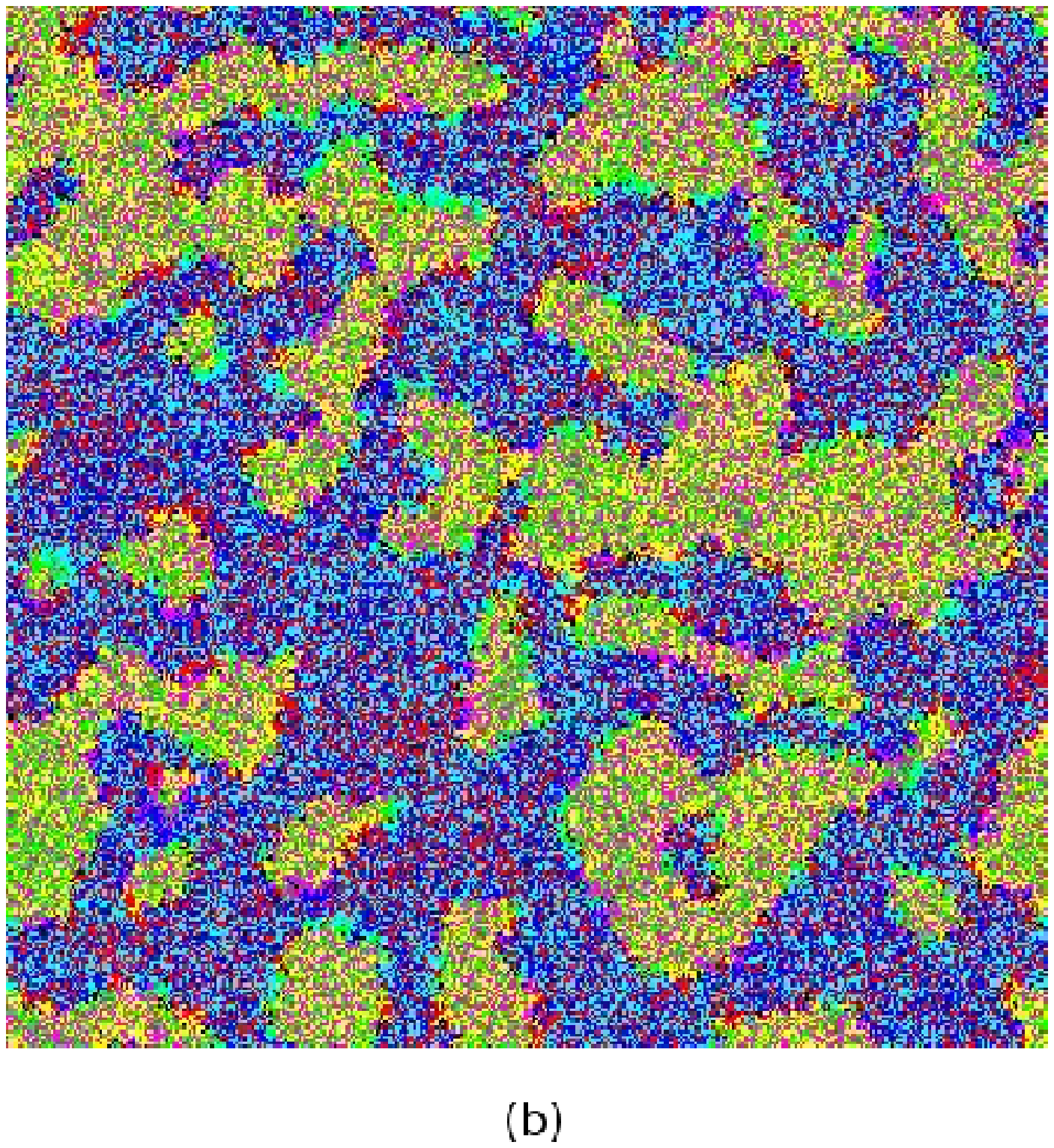}~~
\caption{(Color online) In the model $(6,1)$ two teams are competing again each other, each team being
formed by three non-interacting species.
\label{fig6}
}
\end{figure}

Finally, the case (6,1) is the simple generalization of the rock-paper-scissors game to six species. There has
been some recent research effort aimed at fully characterizing the related four species model \cite{Cas10,Dur11,Dur12,Rom12}.
As for the four species model (or for any other model with an even number of species \cite{Zia11}), two teams of mutually neutral
species form for the six species case, as a quick inspection of
\begin{equation}
\underline{B} =
\renewcommand{\arraystretch}{.45}
\left(
\begin{array}{cccccc}
 0 & 0 & 1 & 0 & 0 & 0 \\
 0 & 0 & 0 & 1 & 0 & 0 \\
 0 & 0 & 0 & 0 & 1 & 0 \\
 0 & 0 & 0 & 0 & 0 & 1 \\
 1 & 0 & 0 & 0 & 0 & 0 \\
 0 & 1 & 0 & 0 & 0 & 0
\end{array}
\right)
\end{equation}
reveals. This then leads to a coarsening process where two different types of domains, one containing
the species [1,3,5], the other the species [2,4,6], compete, as shown in Fig. \ref{fig6}.

\subsection{Nine species}
\begin{figure} [h]
\includegraphics[width=0.48\columnwidth]{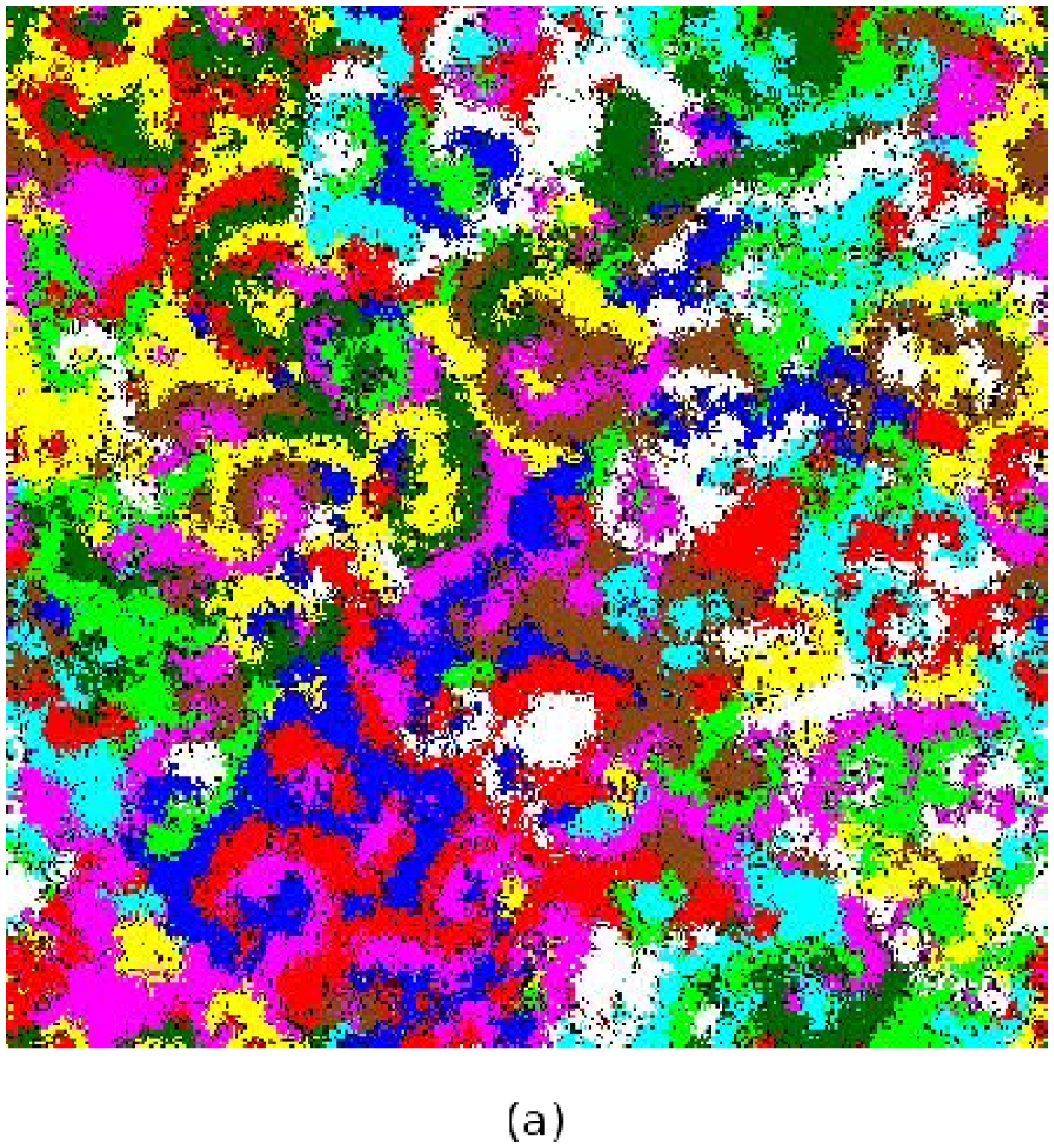}~~
\includegraphics[width=0.48\columnwidth]{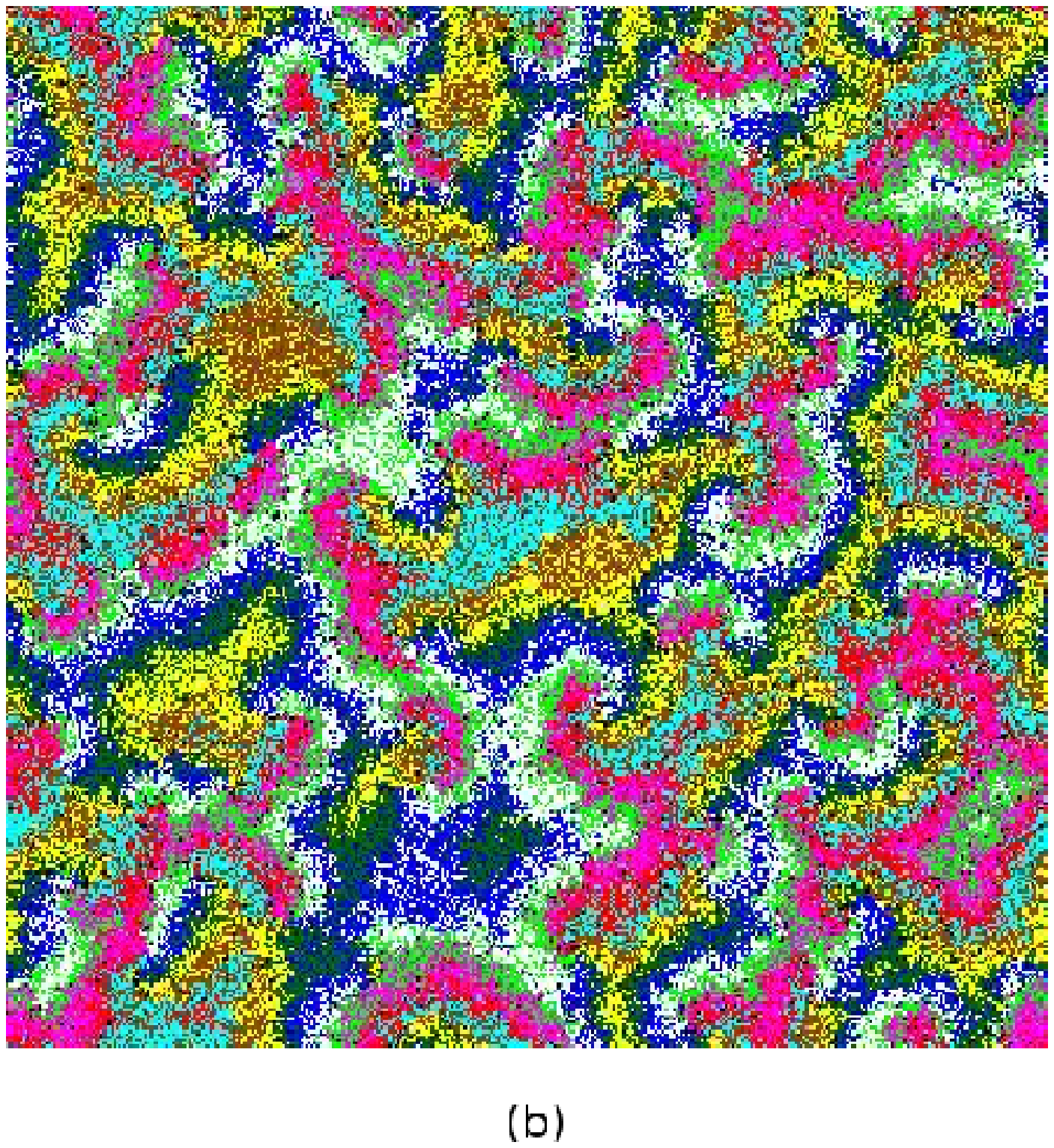}~~
\caption{(Color online) In both cases (9,4) (left) and (9,3) (right) spirals and propagating wave fronts are observed.
Whereas for (9,4) the different waves are formed by a single species, for (9,3) in every wave 
mutually neutral species partially mix.
\label{fig7}
}
\end{figure}

Before considering the general case, we first discuss two situations involving nine species (see
also \cite{Sza07c} for a discussion of a related nine-species system).
As shown in Fig. \ref{fig7} very different looking spirals and wave fronts are encountered
for the two cases (9,4) and (9,3). Whereas in (9,4) the spirals are pure, i.e. formed by
a single species, the waves in (9,3) are characterized by the partial mixing of mutually neutral species.

For the (9,4) case the square of the adjacency matrix reads
\begin{equation}
\underline{B} =
\renewcommand{\arraystretch}{.45}
 \left(
\begin{array}{ccccccccc}
 0 & 0 & 1 & 2 & 3 & 4 & 3 & 2 & 1 \\
 1 & 0 & 0 & 1 & 2 & 3 & 4 & 3 & 2 \\
 2 & 1 & 0 & 0 & 1 & 2 & 3 & 4 & 3 \\
 3 & 2 & 1 & 0 & 0 & 1 & 2 & 3 & 4 \\
 4 & 3 & 2 & 1 & 0 & 0 & 1 & 2 & 3 \\
 3 & 4 & 3 & 2 & 1 & 0 & 0 & 1 & 2 \\
 2 & 3 & 4 & 3 & 2 & 1 & 0 & 0 & 1 \\
 1 & 2 & 3 & 4 & 3 & 2 & 1 & 0 & 0 \\
 0 & 1 & 2 & 3 & 4 & 3 & 2 & 1 & 0
\end{array}
\right)
\end{equation}
from which we immediately obtain that $1 \to 5 \to 9 \to 4 \to 8 \to 3 \to 7 \to 2 \to 6 \to 1$, where $\to$ indicates a preferred relationship
that fixes the order in which the wave fronts propagate through the system. In fact, in our opinion it is one of the most important features
of the matrix $\underline{B}$ that it allows us to immediately read off the order of wave fronts, especially for the more complicated cases.

For the case (9,3) we have that
\begin{equation}
\underline{B} =
\renewcommand{\arraystretch}{.45}
 \left(
\begin{array}{ccccccccc}
 0 & 0 & 1 & 2 & 3 & 2 & 1 & 0 & 0 \\
 0 & 0 & 0 & 1 & 2 & 3 & 2 & 1 & 0 \\
 0 & 0 & 0 & 0 & 1 & 2 & 3 & 2 & 1 \\
 1 & 0 & 0 & 0 & 0 & 1 & 2 & 3 & 2 \\
 2 & 1 & 0 & 0 & 0 & 0 & 1 & 2 & 3 \\
 3 & 2 & 1 & 0 & 0 & 0 & 0 & 1 & 2 \\
 2 & 3 & 2 & 1 & 0 & 0 & 0 & 0 & 1 \\
 1 & 2 & 3 & 2 & 1 & 0 & 0 & 0 & 0 \\
 0 & 1 & 2 & 3 & 2 & 1 & 0 & 0 & 0
\end{array}
\right)
\end{equation}
which yields the relationships $1 \to 6 \to 2 \to 7 \to 3 \to 8 \to 4 \to 9 \to 5 \to 1$. What is special here is that every species tries
to be involved in relationships with the two species with which it is not interacting. As an example consider species 1. From $5 \to 1 \to 6$
follows that species 1 wants to ally with 6, whereas at the same time 5 also wants to ally with 1. 
Species 1 and 6 being mutually neutral, they tend to mix.
A similar behavior takes place between 1 and 5. However, 6 being a prey of 5, this seems to be a bad arrangement for 6 as individuals from species 
5 could simply switch places with those of species 1 in order to prey on 6. Species 6 avoids this by partially mixing with 
species 2 which is a predator
of 5. This complicated game of hide-and-seek results in the fuzzy looking wave fronts shown in Fig. \ref{fig7} for the (9,3) case.

\subsection{General case}
Many of the very rich dynamical features encountered in this family of models are independent of the number $N$ of species.
We here compile these generic situations, characterized by the square of the adjacency matrix.
\begin{itemize}
\item For $(N,N-1)$ models each species forms clusters consisting only of itself. The resulting coarsening process involves domains of $N$
different types, similarly to what is observed in the $N$ states Potts model.
\item The $(N,1)$ cases are the $N$-species generalizations of the usual rock-paper-scissors games where in a cyclic way every species
is interacting with two other species, one being the prey, the other the predator. If $N$ is even, this results in the formation of two teams
composed of mutually neutral species that compete against each other through a coarsening process. If $N$ is odd, spirals are formed. For $N > 3$
these spirals are fuzzy as mutually neutral species partially mix, similar to what we encountered for the (9,3) case, see Fig. \ref{fig7}.
\item If $N-1-\mbox{ceiling}(N/2) \geq r \geq 1$, where $\mbox{ceiling}(N/2)$ is the smallest integer greater or equal $N/2$, 
then two different situations are possible.
If $\mbox{GCD}(r + 1 ,N) \neq 1$, where GCD is the greatest common divisor, then coarsening takes place where $\mbox{GCD}(r+1,N)$ teams composed of
$N/\mbox{GCD}(r+1,N)$ mutually neutral species compete. The model (6,2) is a case where these
conditions are fulfilled, see Fig. 5. If, on the other hand, $\mbox{GCD}(r + 1, N)=1$,
then spirals are formed. As some species do not interact with each other, the resulting
wave fronts are fuzzy, due to the partial mixing of mutually neutral species, see the
case (9,3) discussed above.
\item Finally, if $N-1 > r > N-1-\mbox{ceiling}(N/2)$, we again have to distinguish two different cases. 
If $\mbox{GCD}(r+1,N) \neq 1$, then we have the hybrid case where coarsening takes place
with domains characterized by non-trivial dynamics, see (6,3) for an example. The
number of teams is thereby $\mbox{GCD}(r+1,N)$, with every team being composed of $N/\mbox{GCD}(r+1,N)$
interacting species. If $\mbox{GCD}(r+1, N) = 1$, then spirals are formed with
compact wave fronts that only contain one species, see the cases (9,4) and (6,4).

\end{itemize}

\section{Temporal Fourier analysis of the particle density}

For a more quantitative discussion we focus in this Section on the temporal evolution of the particle density 
for four different cases with eight species, namely (8,1), (8,2), (8,5), and (8,6). The fate of these systems immediately
follows from the rules given in Section IIIC. Whereas for (8,1) we see the formation of two teams, each composed of four mutually
neutral species, that compete against each other, for (8,2) we encounter spirals with fuzzy wave fronts due to the partial mixing
of individuals from different species, similar to what is found for the (9,3) case discussed in Section IIIB. In the (8,5) system
one again encounters two alliances, but this time the members of each alliance are not mutually neutral but prey on each other, 
thus yielding the
previously described complicated coarsening process with spirals inside the coarsening domains. Finally, the (8,6) case results in spirals
with compact wave fronts where every species preferentially hunts the prey that is not at the same time one of its predators.

For the following discussion we fix the values of the
species independent rates at $\delta = \gamma = 0.3$ and $\beta = \alpha = 0.21$.
All runs were done in systems with $256 \times 256$ sites.

\begin{figure} [h]
\includegraphics[width=0.90\columnwidth]{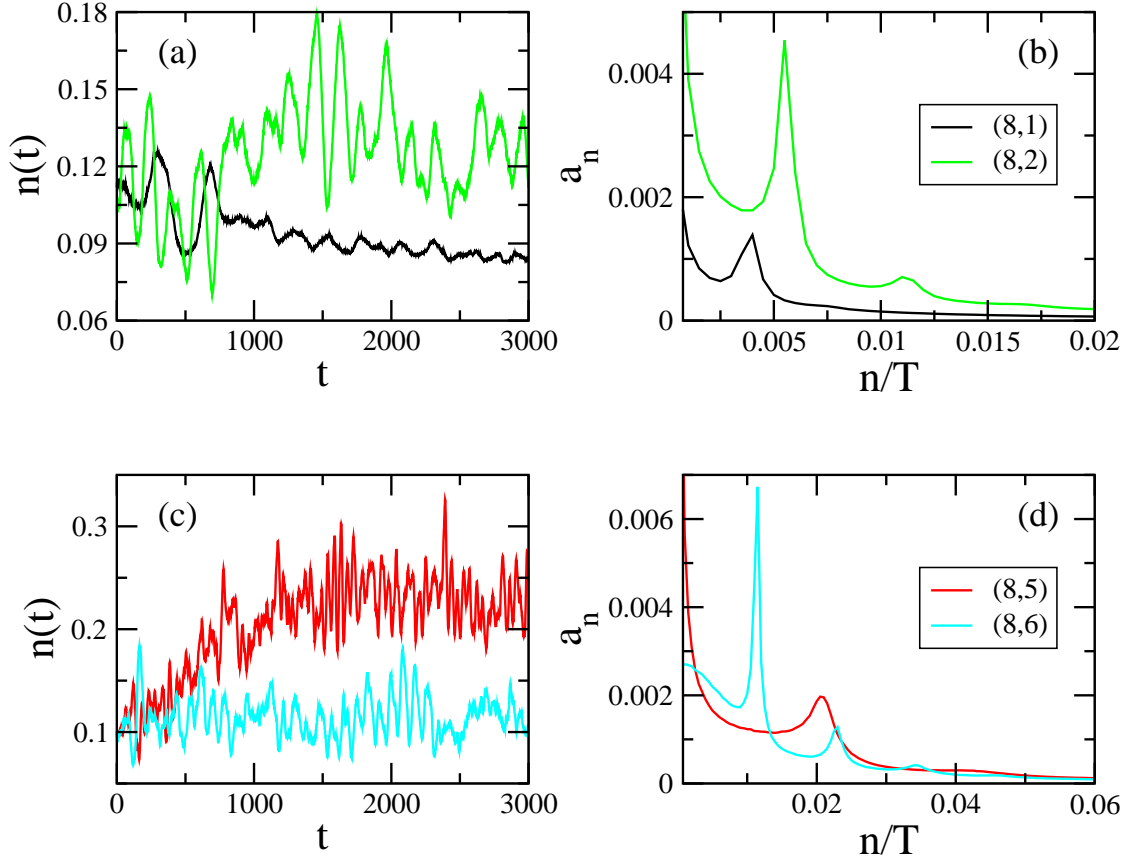}
\caption{(Color online) Time-dependent particle density and the corresponding averaged Fourier transform for (a,b) the models (8,1) and (8,2)
and (c,d) the models (8,5) and (8,6). The systems have $256 \times 256$ sites, and species independent rates $\delta = \gamma = 0.3$
and $\beta = \alpha = 0.21$ are used. The Fourier transform data result from averaging over 3000 independent runs.
\label{fig8}
}
\end{figure}

The temporal evolution of the particle density of one randomly chosen species
is shown in Fig. \ref{fig8}a and \ref{fig8}c for the four different cases. The data displayed in these two panels result
from a single run and are not averaged over an ensemble of simulations. The probability for a certain species to occupy a 
given site at the beginning of the run is 1/9, due to the presence of empty sites. Fig. \ref{fig8}b and \ref{fig8}d 
display the average Fourier transform of the species densities for our four cases. For every run and every species we
compute the Fourier transform
\begin{equation}
\hat{X}_{j,n}=\frac{1}{T}\sum_{t=t_{min}}^{t_{max}}X_{j,t} e^{2\pi i n \frac{t}{T}}
\end{equation}
where $T = t_{max} - t_{min}$ is the chosen time interval and $X_{j,t}$ is the density of species $j$ at time step $t$. This quantity is then
averaged over all eight species as well as over 3000 independent runs. The amplitude of the resulting quantity, $a_n$, is plotted
in Fig. \ref{fig8} as a function of $n/T$, 
where we set $t_{max}=3000$ and $t_{min} = 1000$. 

During the coarsening process that takes place in the (8,1) case the particle density
is characterized by rather regular oscillations with a large period.
This results in a low frequency peak in the Fourier transform at $n/T \approx 0.0037$, with a very weak higher harmonic peak 
at $n/T \approx 0.0075$, see Fig.\ \ref{fig8}b. The finite width of this characteristic peak indicates that the oscillations
will decay and cease after a finite relaxation time.
A rather similar behavior, even so characterized by a much shorter period, is encountered for the coarsening process in the
(8,5) case, see the red curves in Fig.\ \ref{fig8}c and \ref{fig8}d, and this despite the fact that within the coarsening
domains a complicated dynamics persists between the members of a given alliance.
As for the (8,1) and (8,5) cases the coarsening processes take place in a finite system, one of the two competing teams will
ultimately prevail. For the (8,1) run shown in Fig. \ref{fig8}a the population density of the chosen species decreases,
indicating that the team to which this species belongs is losing against the competing team.
This is different for the species chosen to illustrate the (8,5) case in Fig.\ \ref{fig8}c, which increases during the shown time
interval from the initial value 1/9 to much larger values, thereby revealing that the corresponding team has taken over a large
part of the system.

The different types of wave fronts in the systems (8,2) and (8,6) yield complex population oscillations where periods
of rather regular oscillations are separated by bursts of very fast oscillations with small amplitudes, see the green respectively
blue line in Fig. \ref{fig8}a respectively Fig. \ref{fig8}c. The Fourier transforms are characterized not only by very strong
peaks at a characteristic frequency, but multiple higher harmonics are also encountered in these spectra, see the corresponding
curves in Fig. \ref{fig8}b and \ref{fig8}d. Comparing the Fourier transforms for these two cases, one notes that the presence
of compact wave fronts yields a shoulder like feature for very small frequencies, see the blue curve in Fig. \ref{fig8}d. Interestingly,
indications of a similar feature can also be found in the Fourier transform of the original three species May-Leonard model,
see Fig. 2b and 3b in \cite{He11}, which also exhibits compact wave fronts. However, for this three species model, which is the case
(3,1) in our notation, the oscillations are very regular and no higher harmonics are easily identified \cite{He11}.

\begin{figure} [h]
\includegraphics[width=0.90\columnwidth]{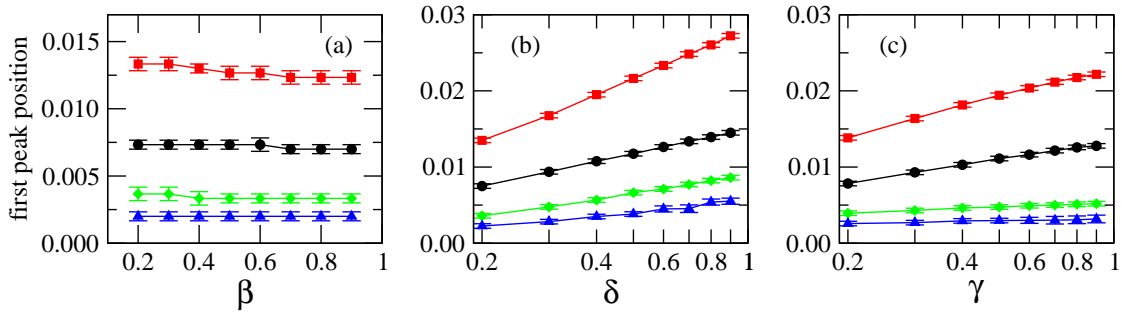}
\caption{(Color online) Position of the characteristic peak of the Fourier transform of the population
density when changing the different rates: (a) changing the diffusion rate $\beta$ while keeping all other rates fixed at 0.2;
(b) changing the predation rate $\delta$ (and, concomitantly, the swapping rate $\alpha = 0.2 (1 - \delta)$) while
keeping the other rates fixed at 0.2; (c) changing the birth rate $\gamma$ (and, concomitantly, the diffusion rate $\beta
= 0.2 (1 - \gamma)$) while
keeping the other rates fixed at 0.2. 
The different symbols correspond to the different cases studied in this Section, namely
blue triangles: case (8,1), green diamonds: case (8,2), red squares: case (8,5), and black circles: case (8,6).
In the panels (b) and (c) the data are plotted in a log-linear fashion, thus
highlighting that the location of the peak depends logarithmically on the rates $\delta$ and $\gamma$.
\label{fig9}
}
\end{figure}

From our discussion follows that the particle densities always display oscillations with a main, characteristic frequency
given by the position of the dominating peak in the Fourier transform. As shown in Fig. \ref{fig9}a, the peak's position
does not exhibit a strong dependence on the mobility of the particles: increasing the swapping rate (the same result is
obtained when increasing the diffusion rate) while keeping the predation and birth rates constant yields for all studied cases
only very minor changes in the spectra. This is different when changing the predation or birth rates, see Fig. \ref{fig9}b
and \ref{fig9}c. Indeed, we find in these cases that the position of the prominent peak displays a  logarithmical dependence
for not too small values of the rates, as illustrated by the resulting straight parts in the log-linear plots.
At present we do not have a good explanation for this observation.

\section{Different coarsening processes}

One of the most intriguing aspects of our family of models comes from the fact that we encounter three very different
types of coarsening processes: (1) the coarsening of pure, compact domains formed by each of the $N$ species, as found
for example in the case (6,5) shown in Fig.\ \ref{fig2}; (2) the coarsening of domains formed by different teams
composed of mutually neutral species, see the case (6,2) in Fig. \ref{fig5} for an example; and (3) the coarsening
of domains formed by alliances of interacting species, so that within the coarsening domains a non-trivial 
dynamics takes place, as it is for example the case for the model (6,3) shown in Fig. \ref{fig4}. In the following we discuss
some aspects of these different types of coarsening processes. 

\subsection{Growth law}

Coarsening systems are characterized by the time-dependent average domain size. In cases where the domains contain
multiple species, which might undergo complicated interactions on their own, we are not dealing with compact clusters
whose size can be easily determined. Instead we proceed as in \cite{Rom12} and determine the typical length in the system
through a space- and time-dependent correlation function. Note that the length determined in that way will follow the
same growth law as the average domain size.

\begin{figure} [h]
\includegraphics[width=0.70\columnwidth]{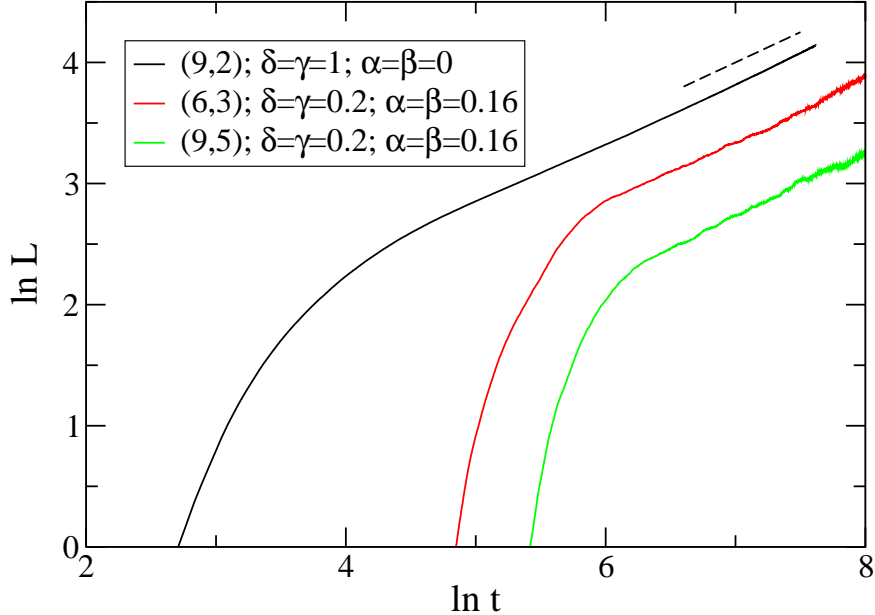}
\caption{(Color online) Time-dependent correlation length for different cases, see the legend. We always
obtain that the long time behavior is given by $L(t) \sim t^{1/2}$, as indicated by the dashed line. These data
result from averaging over 3000 independent simulations for systems composed of $256 \times 256$ sites.
\label{fig10}
}
\end{figure}

We computed the following correlation function
\begin{equation}
C(t, \vec{r}) = \sum\limits_i \left[ \left< m_i(t,\vec{r}) m_i(0,0) \right> - 
\left< m_i(t,\vec{r}) \right> \left< m_i(0,0)
\right> \right]
\end{equation}
where the occupation number $m_i(t,\vec{r})$ equals one if at time $t$ at position $\vec{r}$ a particle of species $i$
is encountered. Otherwise, $m_i(t,\vec{r})=0$. We thereby average over an ensemble of runs characterized by
different initial conditions and different realizations of the noise. The length $L(t)$ shown in Fig. \ref{fig10}
for different cases is then defined as the distance at which the correlation function drops to one third of the 
value it has at $\vec{r} = 0$. 

We show in Fig. \ref{fig10} cases that correspond to the two most interesting types of coarsening where different
teams formed by multiple species are competing against each other. Whereas in the (9,2) case we are dealing with
three teams each composed of three mutually neutral species, in the (6,3) and (9,5) cases two respectively three alliances,
each composed of three interacting species, are in competition. For all cases the correlation length varies as a 
square root of time in the long time limit, as expected for a curvature driven coarsening process. Our result
for the (9,2) case is in agreement with the result obtained previously for the (4,1) model, where two teams composed each of two
mutually neutral partners compete \cite{Rom12,Ave12b}. The same growth law $L(t) \sim t^{1/2}$ has also been observed in \cite{Ave12b} for 
systems with four and five species were the interactions yield the formation of compact single species domains.
The main result of this Section is therefore that also in cases with complicated dynamics inside the
coarsening domains the square root growth law, with corresponds to a dynamic exponent $z=2$, prevails.

\subsection{Interface fluctuations}

In coarsening systems the most complex dynamical behavior is encountered at the interfaces separating different
domains. For the two-dimensional Ising model the interface fluctuations are
known \cite{Abr89} to belong to the so-called Edwards-Wilkinson universality class \cite{Edw82} where after an early time regime the
width of the interface, $W$, increases as $W(t) \sim t^{1/4}$, whereas the equilibrium width increases with the
interface length $L$ as $W_{eq} \sim L^{1/2}$. It is {\it a priori} unclear whether for the multi-species models
studied in this paper the interfaces separating different competing teams belong to the same universality class. In
\cite{Rom12} we provided a first answer to that question through of study of the interface fluctuation of the simple case
(4,1) where the coarsening domains are formed by teams of two mutually neutral species. Indeed, we showed in that
work that we recover the same power-laws as for the two-dimensional Ising model, thereby demonstrating that these
interface fluctuations also belong to the Edwards-Wilkinson university class. The same result is expected for every
case where the team members are not interacting.

However, it is not obvious whether the same conclusions can be drawn for the more complex cases where inside the domains 
predation takes place between the members of a team. This is the question we want to address in the following through
an investigation of the (6,3) model. We thereby focus on a version without empty sites given by the reactions
(\ref{model_singlesite1}) and (\ref{model_singlesite2}), i.e. predation and birth are taking place simultaneously and
mobility is exclusively realized through the swapping of particles sitting on neighboring sites (note that the spirals
seen in Fig. \ref{fig4} are replaced by patches in the absence of empty sites).
For this version of the model we can immediately adapt the approach used in \cite{Rom12} for identifying the location
of the interface.

\begin{figure} [h]
\includegraphics[width=0.80\columnwidth]{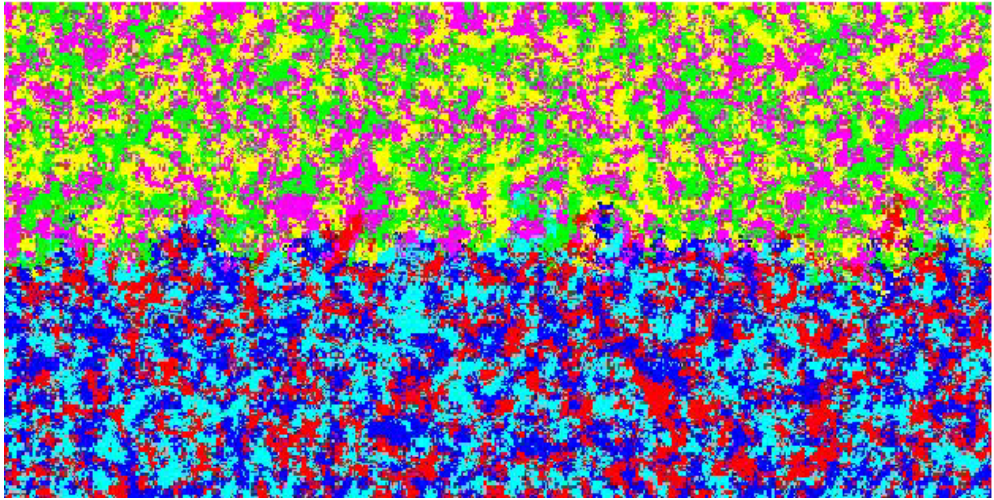}\\[0.5cm]
\includegraphics[width=0.80\columnwidth]{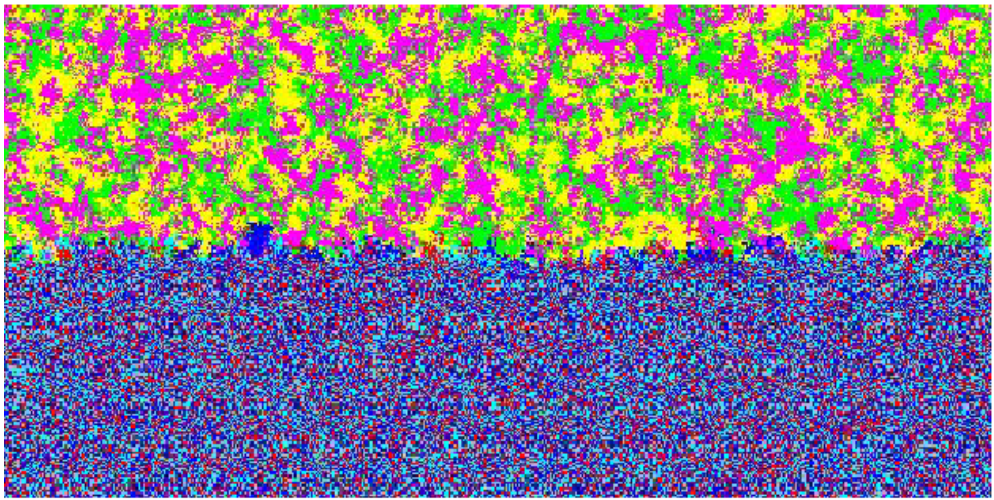}
\caption{(Color online) Interface fluctuations for two versions of the (6,3) model without empty sites,
see Eqs. (\ref{model_singlesite1}) and (\ref{model_singlesite2}). Initially, the two alliances occupy different
halves of the system. Upper panel:  species independent rates $\alpha = 0.2$ and $\delta = 0.8$, yielding 
a complicated dynamics among the members of each alliance. Lower panel: same interactions as for the upper panel, 
with the exception that the predation among members of the alliance in the lower half is not allowed,
thus yielding a team of mutually neutral partners in that half. For the alliance in the upper half the rates are
$\alpha = 0.22$ and $\delta = 0.78$, whereas they are $\alpha = 0.2$ and $\delta = 0.8$ for the team
in the lower half. Both snapshots have been taken after $t=200$ time steps,
for a system containing $600 \times 300$ sites.
\label{fig11}
}
\end{figure}

In order to study interface fluctuations we start with a straight interface separating the two alliances. As already
mentioned, for the (6,3) case every team consists of three interacting species. We prepare the system by
populating each half of the system with one team, where every lattice site can be occupied with the same
probability by any of the three species forming the team. We thereby use periodic boundary conditions along the
interface, whereas perpendicular to the interface reflective boundaries are used. 
Having prepared the system in this way, we then update the
system using the rules (\ref{model_singlesite1}) and (\ref{model_singlesite2}). A snapshot of a system of length
$L =600$ and height $H=300$ taken after 200 MCS is shown in the upper panel of Fig. \ref{fig11}. It is clear that
the initial flat interface roughens due to the competition between the different species. 

In order to determine the local position of the interface we first assign the value $+1$ to all members of one alliance
and the value $-1$ to all members of the other one. This then allows us to determine the interface using approaches initially
developed for the Ising model \cite{DeV05,Mul96} (see also \cite{Rom12}), where for each column $j$ the value $\ell$ is determined that
minimizes the expression
\begin{equation}
v(\ell) = \sum\limits_{k=1}^H \left[ S_{j,k} - p(k-\ell) \right]^2~.
\label{eq}
\end{equation}
The spin $S_{j,k} = \pm 1$ indicates which team can be found at site $(j,k)$. The step function $p(u)$ is defined as
$p = -1$ for $u > 0$ and $p =1$ for $u < 0$. Proceeding in this way, $\ell$ gives the local position of the interface at column $j$.
From this interface profile we then compute the quantity of interest, namely the interface width:
\begin{equation}
W(t) = \sqrt{ \frac{1}{L} \sum\limits_{j=1}^L \left( \ell(j) -  \overline{\ell} \right)^2}~.
\end{equation}
Here $\overline{\ell} = \frac{1}{L} \sum\limits_{j=1}^L \ell(j)$ is the mean position of the fluctuating
interface at time $t$.

\begin{figure} [h]
\includegraphics[width=0.70\columnwidth]{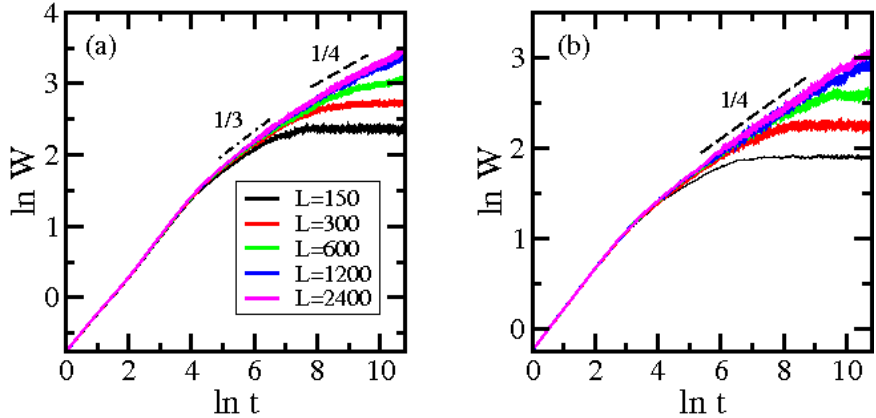}
\caption{(Color online) Interface width as a function of time for systems of different lengths $L$. Panel (a) respectively
(b) corresponds to the upper respectively lower panel in Fig. \ref{fig11}. The data result from averaging over 1000
independent runs. $L$ increases from bottom to top.
\label{fig12}
}
\end{figure}

Fig. \ref{fig12}a summarizes the data we obtain in this way for the (6,3) case. After the initial increase that corresponds
to uncorrelated fluctuations, a first correlated regime sets in that is characterized by an exponent 1/3. This regime, which
is not observed for the Ising model or the (4,1) model \cite{Rom12}, indicates that the complicated interplay between the
different species at the interface yields initial correlated fluctuations characterized by the growth exponent of the
Kardar-Parisi-Zhang universality class \cite{Kar86}. This regime, however, is only transient, as it is followed by a long-time
regime where $W(t) \sim t^{1/4}$, in agreement with what has been found for the simpler (4,1) case. In addition, we find that the equilibrium
interface width scales with the system length as $W_{eq}(L) \sim L^{1/2}$, so that these two exponents agree with the expected
values for the Edwards-Wilkinson universality class.

We also studied a version of the (6,3) model where for the team in the lower half no predation among team members was allowed,
yielding a team of mutually neutral partners, see the lower panel in Fig. \ref{fig11}. Analyzing the interface width for that case,
see Fig. \ref{fig12}b, we find that the transient regime is absent for this simplified dynamics and that the Edwards-Wilkinson
regime immediately sets in after the uncorrelated early time regime. 

This difference in behavior can be rationalized by noting that for the full model patches of a given species enter into the part of the
system that is occupied by the other alliance. 
As our species preys on two of the species in the enemy team, it usually finds many preys in the form
of patches composed by these two species. However, as soon as it encounters a patch of the only enemy species upon which it does not prey,
our patch will disappear rapidly as it is a prey of this third species. This repeating process yields large correlated fluctuations which
are mostly absent for the simplified model where no large patches are formed in the lower half, see Fig. \ref{fig11}. 

\section{Discussion and conclusion}
In spatial systems and in the presence of noise
biodiversity and species coexistence often go hand in hand with complex spatio-temporal patterns. This relationship has
been studied extensively in recent years for population models composed of a few (typically two or three) species. 
Well known examples include stochastic two-species Lotka-Volterra (predator-prey) 
models \cite{Kan05,Mob06,Was07} or stochastic versions of the three-species
spatial May-Leonard model \cite{Rei07}. 

Although important lessons can be learned from these systems with very few species, they are oversimplified as real ecosystems
are not composed by such a small number of species that do not interact with the outside world. On the other hand, increasing
substantially the number of species in order to describe more realistic food webs or ecological networks \cite{Laf10}
very rapidly yields extremely complex systems that can not be
studied in the same systematic way as the few species models.

One promising avenue, intermediate between the simplest models and the large food webs, is to study systems composed of many
species that interact in very specific ways among each other. This is the approach taken in this paper (see also
\cite{Ave12a,Ave12b} for other recent examples). We thereby focus on systems composed of $N$ species, where every species preys
on $r$ other species in a cyclic way. In this way we recover the $N$-species cyclic Lotka-Volterra models as special cases.

Our study reveals a richness of pattern formation not discussed in previous publications. Depending on the values of $N$ and $r$,
various scenarios are encountered. For example, different types of space-time patterns in the form of spirals can emerge.
These spirals can be compact, i.e. formed by a single species, or they can be fuzzy, due to the interpenetration of
members belonging to different species.
There are also three fundamentally different coarsening processes that can be realized.
Whereas in the first one, which is encountered when every species preys on every other species, pure domains are formed that contain
only members of a single species, in a second case we end up with different teams, composed of mutually neutral species, that
compete against each other. The third possibility is a very intriguing one, as here alliances are formed, where the different
members of the alliances are not mutually neutral, but are involved in a predator-prey relationship. Consequently, one
observes coarsening domains with  non-trivial spirals forming {\it inside} the domains.

Remarkably, the different scenarios can be reliably predicted by analyzing the square of the adjacency matrix. Indeed, this
matrix not only allows to read off the composition of the different alliances during coarsening, it also allows to understand the order
of the wave fronts when spirals are formed. Based on an analysis of this matrix, we have written down the conditions
that allow to predict the fate of every system that can be realized in the large class of models discussed in this paper.

From a more quantitative point of view, we studied the time-dependent oscillations that show up in the population densities,
and this both for systems exhibiting spirals and for systems undergoing coarsening. For all cases, the Fourier spectrum
is characterized by a rather broad peak, but higher harmonics of the characteristic frequency of the peak location can
also be observed. Interestingly, the characteristic frequency increases logarithmically with the predation and with the
birth rates. Additional important insights should result from a corresponding analysis of time- and space-dependent density profiles, which
we leave for a future study.

Coarsening processes usually give rise to algebraic dependencies of various quantities on time. We first computed 
space-time correlations and extracted from these data the growth law of the characteristic length $L(t)$ in the system. 
For the cases of coarsening of pure domains or of domains composed of teams of mutually neutral partners we found 
a square root behavior of this length as a function of time, thus confirming the presence of this algebraic law in 
systems with a larger number of species than the few four and five species cases discussed in recent publications \cite{Rom12,Ave12b}.
Interestingly, even when the members of an alliance are themselves in a predator-prey relationship, thus yielding complex
dynamics within the coarsening domains, we find an algebraic growth with the same exponent: $L(t) \sim t^{1/2}$.

We also presented a study of the interface fluctuations for the (6,3) case where we have non-trivial dynamics within the
coarsening domains. Using a version of the model without empty sites, we prepared the system in a state where half of the
system is occupied by one alliance, the other alliance occupying the other half of the systems. This setup allowed us
to compute the time and size dependence of the interface width. Whereas the equilibrium fluctuations as well as the
nonequilibrium fluctuations in the long time limit are described by the exponents of the Edwards-Wilkinson universality class,
we found in addition a well defined transient regime where the growth exponent is 1/3, i.e. the value one has
for the KPZ universality class.

The richness of the family of models studied in this paper is very remarkable and calls for a future in-depth
study of the many intriguing features of spatial-temporal spirals as well as of coarsening processes with non-trivial
internal dynamics within the domains. In this paper we have exclusively focused on numerical simulations. Still,
analytical approaches to investigate some of the intriguing features seem possible. We hope to address this and
other issues in future studies.

\begin{acknowledgments}

This work is supported by the US National
Science Foundation through grant DMR-1205309.

\end{acknowledgments}

\end{document}